\documentclass[fleqn,usenatbib]{mnras}

\usepackage{newtxtext,newtxmath}

\usepackage[T1]{fontenc}

\DeclareRobustCommand{\VAN}[3]{#2}
\let\VANthebibliography\thebibliography
\def\thebibliography{\DeclareRobustCommand{\VAN}[3]{##3}\VANthebibliography}

\usepackage{graphicx}	
\usepackage{amsmath}	
\usepackage{xspace}
\usepackage[dvipsnames, svgnames]{xcolor}

\newcommand{\firons}{\ensuremath{f^\star_\mathrm{iron}}\xspace}
\newcommand{\fironp}{\ensuremath{f^\mathrm{c+m}_\mathrm{iron}}\xspace}
\newcommand{\feh}{\ensuremath{\left[\mathrm{Fe/H}\right]}\xspace}
\newcommand{\sih}{\ensuremath{\left[\mathrm{Si/H}\right]}\xspace}
\newcommand{\mgh}{\ensuremath{\left[\mathrm{Mg/H}\right]}\xspace}

\newcommand{\alphafe}{\ensuremath{\left[\mathrm{\alpha/Fe}\right]}\xspace}
\newcommand{\teff}{\ensuremath{T_\mathrm{eff}}\xspace}

\title[Star--Planet Compositional Ties]{No statistically significant evidence for a correlation between stellar and planetary composition}

\author[D. A. Turner et al.]{
Daisy A. Turner$^{1}$\thanks{E-mail: dat936@student.bham.ac.uk},
Annelies Mortier$^{1}$,
James G. Rogers$^{2}$,
Jo Ann Egger$^{3}$,
Alix V. Freckelton$^{1}$,
\newauthor
Tim Lichtenberg$^{4}$,
Anjali A. A. Piette$^{1}$
\\
$^{1}$School of Physics and Astronomy, University of Birmingham, Edgbaston, Birmingham B15 2TT, UK\\
$^{2}$Institute of Astronomy, University of Cambridge, Madingley Road, Cambridge CB3 0HA, UK\\
$^{3}$European Space Agency (ESA), European Space Research and Technology Centre (ESTEC), Keplerlaan 1, 2201 AZ Noordwijk, The Netherlands\\
$^{4}$Kapteyn Astronomical Institute, University of Groningen, 9747 AD Groningen, The Netherlands\\
}

\date{Accepted XXX. Received YYY; in original form ZZZ}

\pubyear{\the\year{}}

\begin{document}
\label{firstpage}
\pagerange{\pageref{firstpage}--\pageref{lastpage}}
\maketitle

\begin{abstract} 

The chemical compositions of planets and their hosts stars are intrinsically linked, having formed from the same protostellar material.
Characterising this relationship provides key constraints on the processes governing planetary formation and evolution.
Focusing on host stars with near-solar chemical abundances limits our understanding of how planetary composition varies across a broader chemical parameter space.
Expanding the sample to include stars with compositions markedly different from the Sun is therefore essential for building a complete picture of star--planet compositional connections.
Here, we focus specifically on iron-poor hosts since they are more likely to be alpha-enhanced and represent some of the most chemically distinct stars relative to the Sun (i.e., thick disc stars).
We present a sample of 45 stars hosting 64 planets across the super-Earth and sub-Neptune regimes, for which we homogeneously obtain new stellar parameters including abundances, re-derive planetary mass and radius, and model the bulk interior compositions of their planetary companions.
No statistically significant evidence for a correlation between stellar and planetary composition was found in our sample.
We suggest that this null result is primarily driven by the large uncertainties inherent to both compositional proxies, which may obscure an underlying relationship.
Quantifying these uncertainties is therefore a critical step that previous studies have not fully addressed.
Furthermore, even with improved precision, uncovering such a relation may require a higher-dimensional treatment that accounts for additional parameters such as planetary equilibrium temperature.

\end{abstract}




\section{Introduction}

Missions such as \textit{TESS} \citep{Ricker2015} and \textit{Kepler} \citep{Borucki2010, Howell2014} provided us with a large pool of small ($R_\mathrm{p}\leq4\,\mathrm{R_\oplus}$) exoplanets, providing a solid statistical base to investigate how planetary properties vary across the population.
One particular area of focus is the extent to which a planet's composition is inherited from its host star, and whether it is therefore predictable from stellar spectral observations alone.
Pinning down any possible compositional connections between stars and the planets they host would shed light on how planets are formed and how they change over time, with stellar chemistry potentially acting as a predictor of planetary interiors \citep{Spaargaren2023ApJ,Putirka2024RvMG}.

The assumption that a star--planet compositional relationship exists is not arbitrary; it is physically motivated by the fact that both bodies form from the same molecular cloud.
The elemental ratios present in stellar photospheres are often used as proxies for the primordial composition of the protoplanetary disc from which planets formed \citep{Thiabaud2015}, making some degree of a relation a natural expectation.
This is evidenced by the Solar System planets: Earth, Venus, and Mars all have Mg/Fe ratios that are comparable to that of the Sun \citep{Behmard2025}.
Extrapolating this to exoplanets, however, is non-trivial.
Unlike the planets in our Solar System, exoplanetary compositions must be inferred from their bulk properties (namely mass and radius), which they themselves are derived relative to their host star.
In practice, this inference for exoplanets relies on interior structure models \citep{Dorn2015AA,Haldemann2024,Egger2024,Nicholls2026,Attia2026}, and obtaining the precise planetary masses and radii that these models require means combining photometric observations with dedicated radial velocity follow-up.
The latter of these two requirements is observationally expensive for small planets, explaining why sample sizes in similar studies \citep{Adibekyan2021, Adibekyan2024, Ross2025} have remained small.
Despite how sophisticated they are, current planetary interior models are still simplified when compared to what is known about the Earth's interior, highlighting that further development of these models is a necessity.

Still, there is an established compositional link between giant planets and their host stars, with one of the earliest strong correlations emerging between hot Jupiter occurrence rates and host star metallicity \citep{Gonzalez1997, Santos2004, Fischer2005ApJ, Mortier2013}.
In contrast, no such correlation was found for small planets \citep{Buchhave2012, Buchhave2014}.
In terms of composition, several studies have found compositional links for giant planets and their host stars \citep{Thorngren2016, Teske2019, Teske2024}.
Recent work suggests that an analogous relationship may hold for smaller, rocky planets \citep{Adibekyan2021, Liu2023, Brinkman2024, Plotnykov2026}.
However, firm conclusions remain elusive for several main reasons.

Firstly, the quoted uncertainties on stellar abundances are frequently underestimated, often falling below the systematic errors introduced by the atmospheric models used to derive them \citep{Tayar2022, Sandford2023}; when propagated through to compositional proxies, this leads to significant underestimates of the true uncertainty on stellar compositions.
Secondly, as aforementioned, sample sizes have largely been restricted to small numbers of super-Earths ($R_\mathrm{p}\lesssim2\,\mathrm{R_\oplus}$), limiting what can be said about the broader small-planet population.

Expanding this work to include sub-Neptunes ($2\lesssim R_\mathrm{p}/\mathrm{R_\oplus}\leq4$) is strongly motivated by the existence of the radius valley, a relative paucity of planets at approximately 1.5--2$\,\mathrm{R_\oplus}$ \citep{Fulton2017, VanEylen2018, Ho2023}, thought to arise from atmospheric escape processes such as photoevaporation or core-powered mass loss \citep{Owen2013, Lopez2013, Ginzburg2018, Gupta2019}.
Planets on either side of this gap have likely undergone distinct evolutionary histories, meaning that comparing their compositions offers a way to disentangle the chemical imprint of formation from that of subsequent evolution \citep{Rogers2025}.

In this paper, we expand existing samples of small planets to include sub-Neptunes and compare stellar composition to that of the planets they host using a variety of compositional proxies.
Section~\ref{sec:st_data} presents the stellar data that was used and
Section~\ref{sec:pl_samp} presents the planetary sample used in this study and details the planetary and stellar constraints applied.
In Section~\ref{sec:st_analysis}, we outline the method by which we obtain stellar abundances and isochronal information.
Section~\ref{sec:link} probes a link between stellar and planetary composition using the planet's bulk parameters while Section~\ref{sec:int} approaches it using interior modelling analyses.
An extensive discussion is presented in Section~\ref{sec:disc}, including parameter precision and accuracy, using the Solar System as a benchmark, a possible relation with equilibrium temperature, and outlining current limitations.
We summarise our conclusions in Section~\ref{sec:conc}.

\section{Stellar Data}\label{sec:st_data}

In this study we used approximately 2\,000 high-resolution optical spectra to measure stellar atmospheric parameters and individual abundances of 45 stars. All spectra are available via public archives, and were taken using a range of instruments: the High Accuracy Radial velocity Planet Searchers (HARPS and HARPS-North), the Echelle SPectrograph for Rocky Exoplanets and Stable Spectroscopic Observations (ESPRESSO), and the Spectrographe pour l'Observation des Ph\'enom\`enes des Int\'erieurs stellaires et des Exoplan\`etes (SOPHIE).
These instruments were selected to maximise the sample size while also ensuring that similar wavelength ranges were covered.
Archival spectra were sourced from the highest-resolution spectrograph available for each target, in order of preference: ESPRESSO, HARPS/HARPS-N, then SOPHIE.
A summary of the programmes from which spectra were taken is available in the acknowledgements; a list of the spectral sources for each target is given in Table~\ref{tab:instruments}.

\subsection{HARPS spectroscopy}

HARPS is a fibre-fed spectrograph installed at the European Southern Observatory (ESO) 3.6\,m telescope in La Silla, Chile \citep{Pepe2000, Mayor2003}, with an average resolving power of $\mathcal{R}=120\,000$ and a wavelength range of 380--690\,nm.

HARPS spectra were available for 21 of our 45 targets, and for these we used all of the available spectra.
All spectra were processed with the standard pipelines and directly downloaded from the ESO archive.

\subsection{HARPS-N spectroscopy}

Located at the 3.6\,m Telescopio Nazionale Galileo (TNG) in La Palma, Spain, HARPS-N is a high-precision, high-resolution, fibre-fed \'echelle spectrograph \citep{Cosentino2012}, that serves as the northern hemisphere's enhanced, optimised version of the original HARPS spectrograph.
It has an average resolving power of $\mathcal{R} = 115\,000$, and covers a wavelength range of 383--693\,nm.

29 of the 45 targets in our sample have HARPS-N available on via the TNG archive; we used all available spectra for these targets.
All spectra were processed with the standard pipelines.

\subsection{ESPRESSO spectroscopy}

The ESPRESSO spectrograph is installed at the incoherent combined Coud\'e focus of the Very Large Telescope (VLT) at Paranal Observatory in Chile \citep{Pepe2014, Pepe2021}.
The full wavelength range for all modes is 380--788\,nm, split into a blue arm (380--525\,nm) and a red arm (525--788\,nm).
The instrument has an average resolving power of $\mathcal{R}=140\,000$ in high-resolution mode (used for majority of targets) and $\mathcal{R}=70\,000$ in medium resolution mode (only used for TOI-402).

We used all available ESPRESSO spectra for our targets (12 of 45).
All spectra were processed with the standard pipelines and directly downloaded from the ESO archive.

\subsection{SOPHIE spectroscopy}

Mounted on the 1.93\,m telescope at the Haute-Provence Observatory in France, SOPHIE is an \'echelle spectrograph with two resolution modes \citep{Perruchot2008} and a wavelength range of 387--694\,nm.
All SOPHIE spectra used in this study were taken in high resolution mode (average resolving power of $\mathcal{R}=75\,000$), instead of the high efficiency mode (average resolving power of $\mathcal{R}=40\,000$).

For the remaining 2 targets, we utilised SOPHIE spectra.
All spectra were processed with the standard pipelines and directly downloaded from the SOPHIE archive.

\subsection{Additional stellar data}
\label{sec:star_extra}

In Section \ref{sec:stelparam}, we re-derive in a homogeneous manner the stellar masses and radii. To accomplish this, we additionally collect from the public archives for all our stars the Gaia DR3 parallax \citep{GaiaCollaboration2023}, and the photometric magnitudes in the following bands: B, V, J, H, K, W1, W2, W3.

\section{Planetary Sample}\label{sec:pl_samp}

\begin{figure}
	\includegraphics[width=0.99\columnwidth]{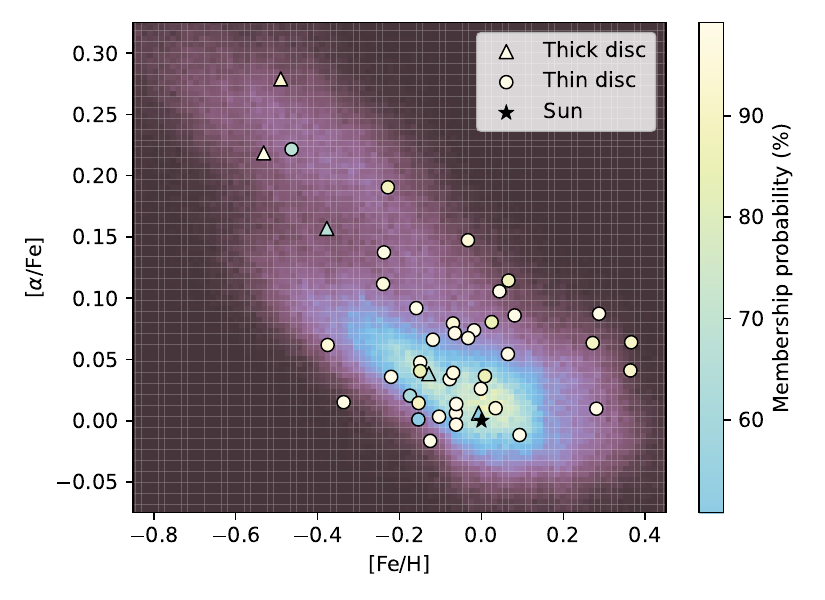}
    \caption{2D histogram showing the systems in this sample. Thick disc and thin disc stars are plotted as triangles and circles, respectively, and the Sun is indicated with a black star. The probability of thick or thin disc membership for each star is shown by the colour of the markers. The colour scale for the background population is representative of number density. Standard quality filters (\texttt{snr\_px\_ccd3>30}, \texttt{flag\_X\_fe==0}, and texttt{flag\_sp==0}) have been applied.}
    \label{fig:gal}
\end{figure}

In order to comprehensively study the link between stellar and exoplanetary composition, it is essential to compare the predominant elements present in planets to those readily measurable in their host stars.
This analysis can be conducted by modelling planetary interiors and contrasting the amounts of their constituent elements with those observed in stellar atmospheres.
Previous studies \citep{Adibekyan2021, Adibekyan2024, Behmard2025, Brinkman2025, Ross2025} have often only included super-Earths ($R_\mathrm{p}\lesssim2\,\mathrm{R_\oplus}$; those below the radius valley), but we aim to include sub-Neptunes ($2\lesssim R_\mathrm{p}/\mathrm{R_\oplus} \lesssim 4$) in our analysis of planetary interiors.
As an approximation, small planets ($R_\mathrm{p}\leq4\,\mathrm{R_\oplus}$) comprise an iron core, a silicate-based mantle, and a fully mixed volatile envelope of water and H/He \citep[e.g.,][]{Rogers2015}.
The relevant stellar abundances, such as iron and the $\alpha$-elements (e.g., silicon, magnesium, oxygen, and titanium), are readily obtainable from stellar spectra.
In turn, the key elements in each of the planetary layers can be quantified from those spectra: iron dominates in the core but is also present within the mantle (as accounted for in the planetary models used here: Section~\ref{sec:int}), while the $\alpha$-elements (e.g., silicon, magnesium, oxygen) are key in the mantle.
Therefore, we require a stellar sample that is diverse in \alphafe space, as probing stars with a range of \alphafe values should allow us to study planets with diverse compositions, provided that such a connection exists.
Fig.~\ref{fig:gal} shows the extensive distribution of the stellar sample across both \alphafe and \feh, with kinematic thin and thick disc members indicated by circles and triangles, respectively.
The background population, published by \citet{Buder2025}, is representative of the Milky Way.

The PlanetS\footnote{\url{https://dace.unige.ch/exoplanets/}} catalogue \citep{Otegi2020, Parc2024} was utilised as the foundation for sample selection as it exclusively includes planets with precisely measured masses and radii, with relative errors of $\sigma_{M_\mathrm{p}}/M_\mathrm{p}\le25\,\%$ and $\sigma_{R_\mathrm{p}}/R_\mathrm{p}\leq8\,\%$.
This precision restriction on the global parameters is necessary for precisely modelling the interiors of planets.

To facilitate robust stellar abundance measurements, we restrict our sample to bright FGK dwarfs ($4000 \leq T_{\mathrm{eff}} \leq 7000\,\mathrm{K}$; $\log g \geq 4$) that are not in known binary or multiple systems.
We further require that each target have an apparent magnitude $m_\mathrm{V}\leq 13$, ensuring that the stars are sufficiently bright to obtain high signal-to-noise (S/N) spectra for precise abundance analysis.
As the stellar spectra were obtained from public archives, targets were excluded if there exist no public spectra that, when co-added, exceed an S/N of approximately 300.

\begin{figure}
	\includegraphics[width=0.99\columnwidth]{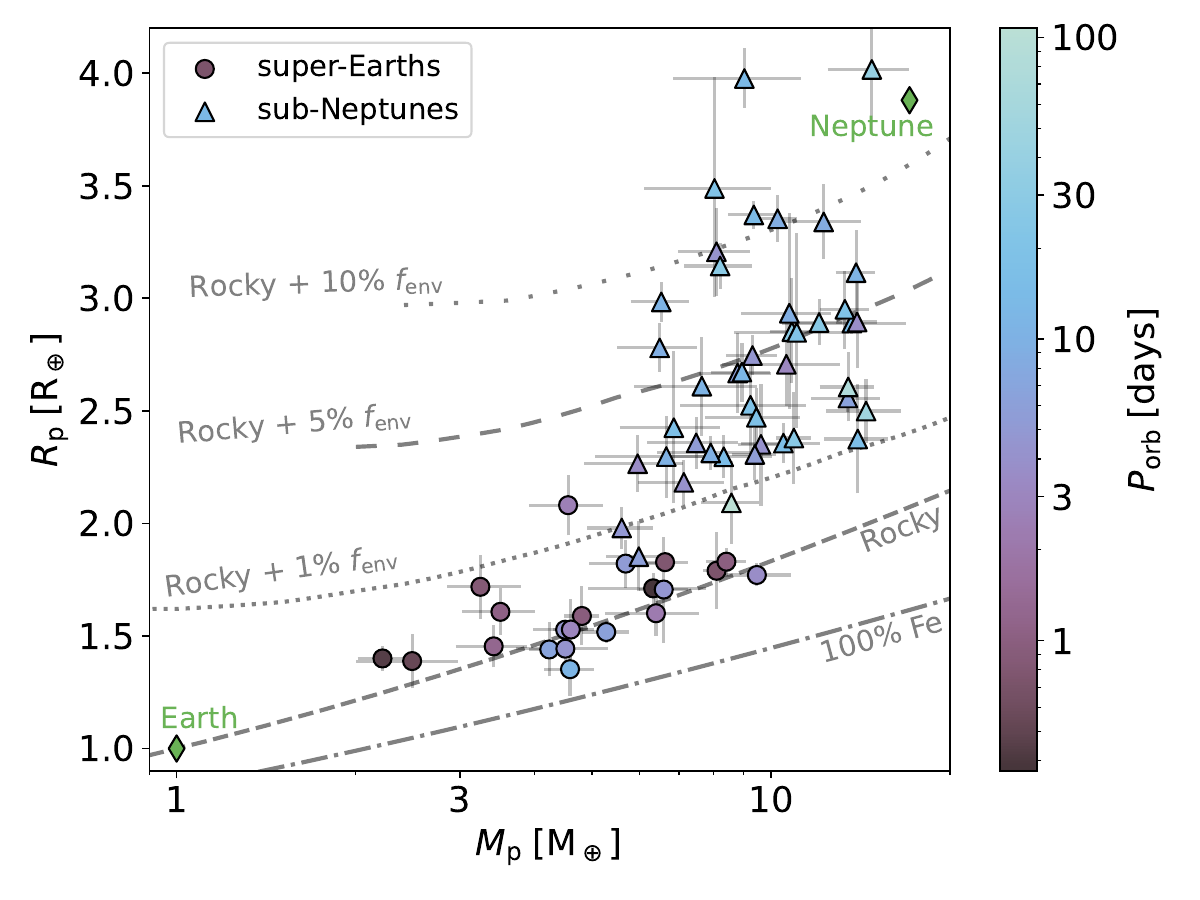}
    \caption{Mass--radius diagram showing the planetary sample that is the focus of this work. Super-Earths are denoted by the circles, and sub-Neptunes by the triangles. Earth and Neptune are indicated by the green diamonds. Compositional lines are from \citet{Zeng2019} for the 100\% iron and rocky, Earth-like lines, and \citet{Chen2016} for the rocky lines with various envelope mass fractions.}
    \label{fig:mr}
\end{figure}

Following the implementation of the criteria above, we retain a sample size of 45 stars with 64 orbiting planets, 21 of which are super-Earths and the remaining 43 being sub-Neptunes.
These are presented in Fig.~\ref{fig:mr}, alongside relevant compositional relations \citep{Chen2016, Zeng2019}, and Earth and Neptune.

\section{Stellar Analysis}\label{sec:st_analysis}

\subsection{Atmospheric analysis}

We followed the ARES+MOOG method, described in, e.g., \citet{Santos2004, Sousa2008, Sousa2011, Santos2013, Tsantaki2013, Mortier2014}, to obtain detailed stellar abundances for all stars in the sample. To facilitate this analysis, we limited the sample to only include FGK host stars. 
The spectra of cooler stars, such as M dwarfs and even late K dwarfs, suffer from blending and the presence of molecular signatures, impeding detailed stellar analysis.
Hotter stars (O, B, and A stars) host fewer planets \citep{Giacalone2025} and are often fast rotators.
This gives rise to spectral broadening, which renders the equivalent width method (i.e., ARES+MOOG) inapplicable.

Given a line list, ARES \citep[Automatic Routine for line Equivalent widths in stellar Spectra, ][]{Sousa2007} measures the equivalent widths of select lines.
MOOG \citep[][2019 version via \texttt{pyMOOGi}\footnote{\url{https://github.com/madamow/pymoogi}}]{Sneden1973} then measures the individual abundances ensuring ionisation equilibrium and Fe\,\textsc{i}--Fe\,\textsc{ii} excitation balance in local thermodynamic equilibrium (LTE).

The base line list for Fe\,\textsc{i} and Fe\,\textsc{ii} was sourced from \citet{Sousa2008}, with line additions and removals made by \citet{Neves2009} and \citet{Adibekyan2012}.
Each star was ran with either the Kurucz ATLAS9 plane-parallel model grid\footnote{\url{http://kurucz.harvard.edu/grids.html}} \citep{Kurucz1993} or the Model Atmospheres with a Radiative and Convective Scheme (MARCS) plane-parallel model grid\footnote{\url{https://marcs.astro.uu.se}} \citep{Gustafsson2008}.
Stars with temperatures $T_\mathrm{eff}\lesssim 4900\,\mathrm{K}$ were ran with the MARCS models and with an adapted iron line list from \citet{Tsantaki2013} as they are optimised for cooler stars.
Oscillator strengths ($\log{gf}$) for all lines were recalculated using the 2019 MOOG version with a HARPS-N solar spectrum and solar reference abundances from \citet{Asplund2009}.

In a first run, using just the iron lines, we obtain the global atmospheric parameters: stellar effective temperature ($T_\mathrm{eff}$), stellar surface gravity ($\log{g}$), microturbulent velocity ($\xi$), and iron abundance ([Fe/H]). Then we use these atmospheric parameters and EWs of several lines of individual elements to measure the individual chemical abundances.

For all stars, we thus obtained abundances for Fe\,\textsc{i}, Mg\,\textsc{i}, Si\,\textsc{i}, and Ti\,\textsc{i}.
The latter three are alpha elements (those produced via the alpha process in stars), and are of particular interest as they constitute the bulk of planetary mantles.
To compare stellar compositions with planetary compositions, we focus on the elements that are most present in both bodies: iron content versus silicate (or alpha element) content. 
We quantify this comparison by using stellar iron mass fraction, \firons, as a proxy for relative stellar composition.
Stellar iron mass fraction is calculated as follows:
\begin{equation}\label{eq:firons}
    \firons = \frac{m_\mathrm{Fe}}{m_\mathrm{Fe} + m_\mathrm{MgSiO_3} + m_\mathrm{MgSiO_4} + m_\mathrm{SiO_2}},
\end{equation}
where $m_i$ is the mass of a molecule, $i$.
We follow the method described by \citet{Adibekyan2021} and the references therein.

\subsection{Isochronal analysis}
\label{sec:stelparam}

With \teff  and [Fe/H] obtained from the atmospheric analysis for all stars, we can use isochrones and stellar tracks to obtain updated and homogeneous stellar masses, radii, and ages.
We follow the methods introduced by \citet{mortier2020}.
The public \textsc{isochrones} software package \citep{morton2015} was used in conjunction with the stellar models from the MESA Isochrones and Stellar Tracks \citep[MIST; ][]{dotter2016}.
Next to the spectroscopic \teff and [Fe/H], we used the Gaia DR3 parallax and the photometric magnitudes mentioned in Section~\ref{sec:star_extra}.

Nested sampling is used to explore the parameter space and calculate the posterior distributions for stellar mass, radius, and age.
The final adopted parameters are the median values from these posterior distribution, with uncertainties representing the 16th and 84th percentiles.
See Fig.~\ref{fig:new_vs_arch} for a comparison between the archival and recalculated stellar parameters.

\section{Probing a star--planet compositional link}
\label{sec:link}

One key shortcoming of previous studies into a compositional link is the lack of homogeneity in both their planetary and stellar parameters.
Planetary mass and radius values drawn from heterogeneous sources will have been derived using inconsistent stellar parameters (different stellar masses and radii); small offsets in stellar parameters introduce scatter in the planetary parameters.
This is particularly consequential for compositional studies: if the methods used to obtain the stellar parameters that underpin the planetary properties are not consistent, any inferred trend may be an artefact of the inhomogeneity rather than a reflection of the influence of the host star's composition.
To address these inhomogeneities, we recalculate the planetary mass based on the updated stellar masses derived in Section \ref{sec:stelparam}, and use the observational parameter $R_\mathrm{p}/R_\star$ as opposed to $R_\mathrm{p}$ on it own.
The planetary orbital periods were taken from the PlanetS database.

\begin{figure}
	\includegraphics[width=0.99\columnwidth]{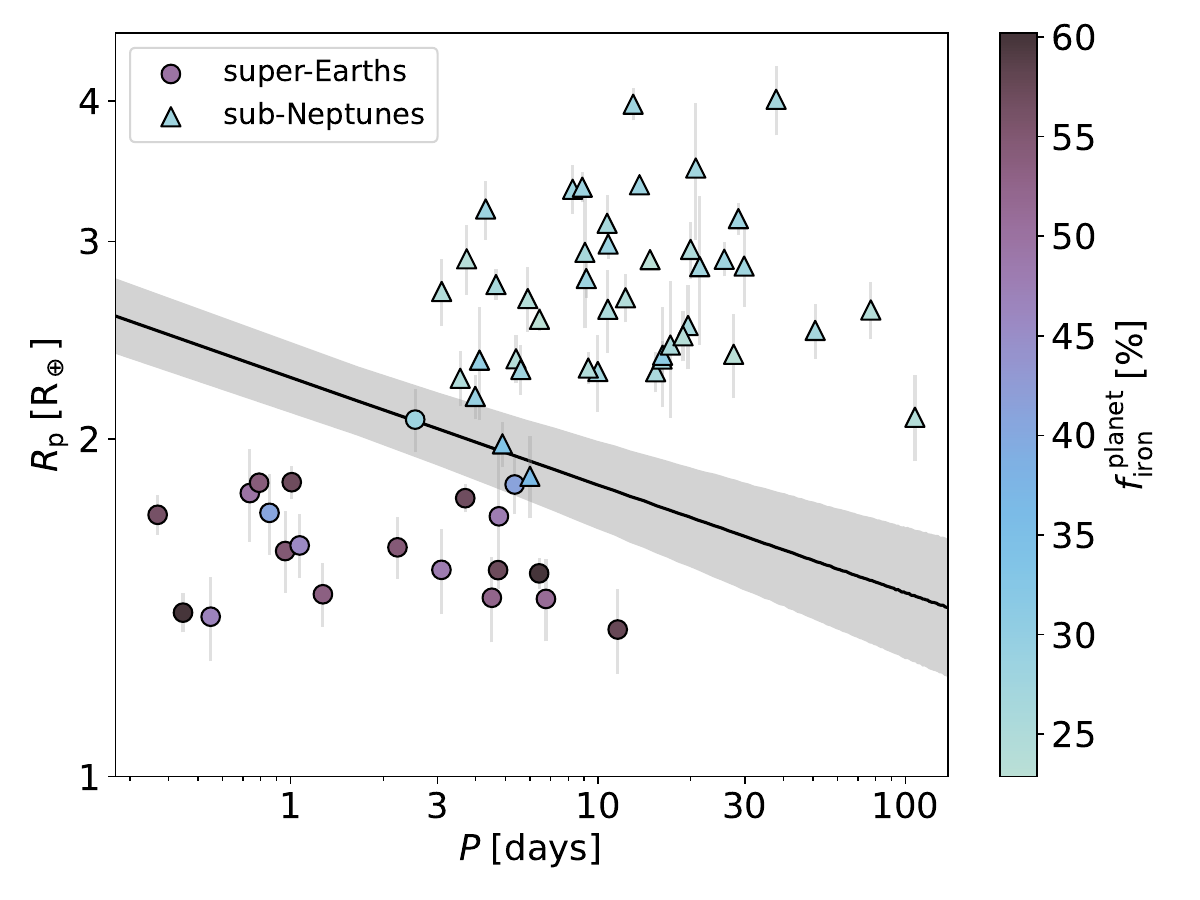}
    \caption{Radius valley plot showing the sub-populations of super-Earths and sub-Neptunes as circles and triangles, respectively. The black line and grey shaded region denote an average radius valley \citep{Ho2023} for the population and its $1\sigma$ uncertainty region, respectively.}
    \label{fig:rad_valley}
\end{figure}

Bulk planetary properties are useful in many contexts, for example, separating planets into sub-populations.
Super-Earths and sub-Neptunes are commonly delineated purely by their radius, and are suggested to occupy the regions where $R_\mathrm{p}\lesssim2\,\mathrm{R_\oplus}$ and $2\lesssim R_\mathrm{p} / \mathrm{R_\oplus}\lesssim4$, respectively.
However, it is best practice to utilise the radius valley \citep{Fulton2017} to separate these two populations, as there exists dependence on orbital period, stellar mass and stellar age.
In this work, we used the stellar-mass-dependent and age-dependent radius valley as described by \citet{Ho2023} to separate super-Earths and sub-Neptunes into two different planetary classes.
This is shown in Fig.~\ref{fig:rad_valley}, where super-Earths and sub-Neptunes are separated by the radius valley.

In addition, one of the main properties we can infer about a planet with a mass and a radius measurement is its bulk density, $\rho_\mathrm{p}\propto M_\mathrm{p}\,R^{-3}_\mathrm{p}$.
This can often serve as a general indicator of the interior structure of a planet.
A high-density planet can be thought to have a larger, more metal-rich core encased within a lighter, more silicate-rich mantle, leading to a higher core mass fraction.
For example, a higher bulk density could imply a relatively larger metal-rich core and/or silicate mantle, while a lower bulk density could point to a higher volatile fraction.

\begin{figure}
	\includegraphics[width=0.99\columnwidth]{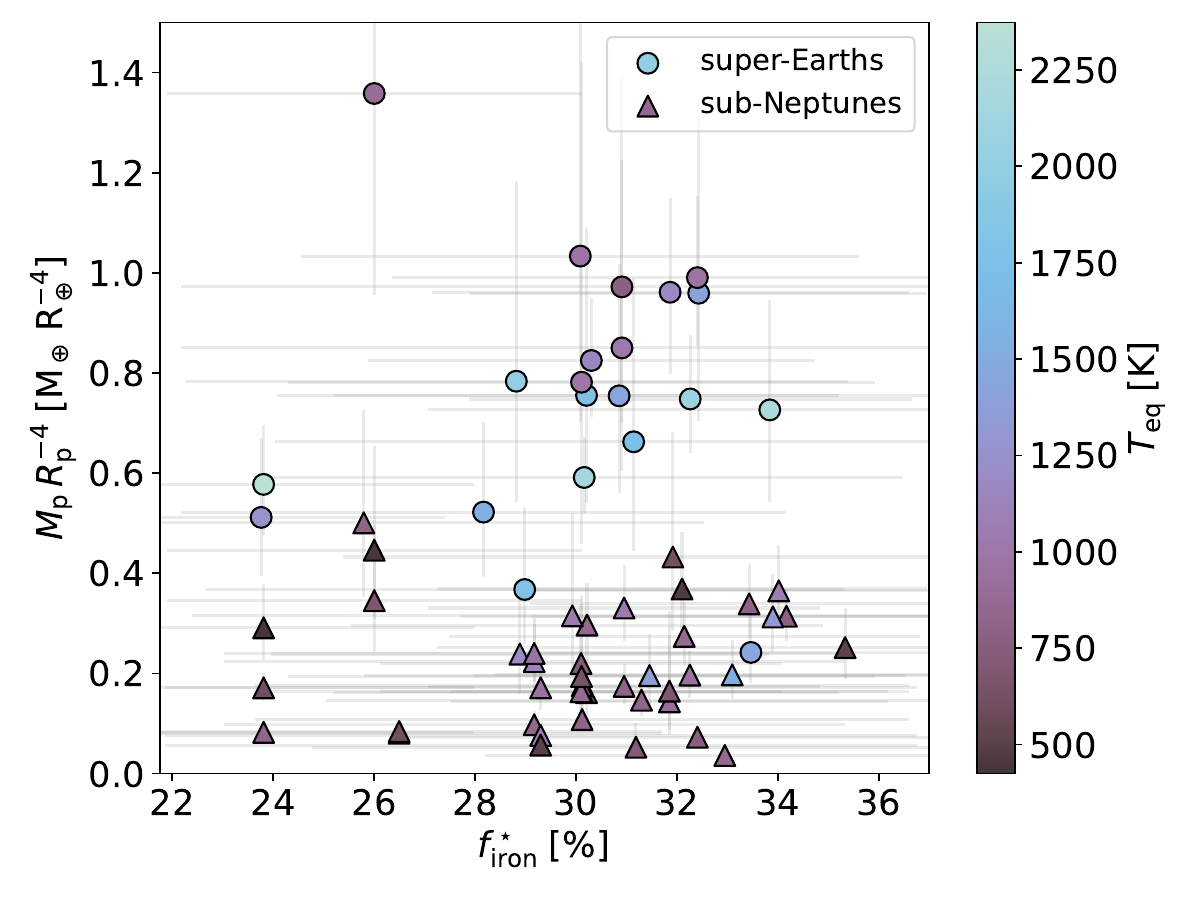}
    \caption{Planetary density scaled by an additional $R_\mathrm{p}^{-1}$ term to account for compression ($M_\mathrm{p}\,R_\mathrm{p}^{-4}$) versus stellar iron mass fraction. Super-Earths and sub-Neptunes are shown as circles and triangles, respectively.}
    \label{fig:firon_mr4}
\end{figure}

However, using planetary bulk density as a metric for composition across a range of planetary classes does not account for internal compression.
Fig.~\ref{fig:firon_mr4} compares $M_\mathrm{p}\,R_\mathrm{p}^{-4}$ to the stellar iron mass fraction, \firons (Equation~\ref{eq:firons}).
Here, $M_\mathrm{p}\,R_\mathrm{p}^{-4}$ is a parameter that is analogous to $\rho_\mathrm{p}$, but better reflects similar composition due to its additional factor of $R_\mathrm{p}^{-1}$ to account for compression \citep{Baumeister2025}.
With the low-\firons region (\firons$\lesssim 28$ per cent) having been further populated, we can see from Fig.~\ref{fig:firon_mr4} that using these metrics, there is no discernable relationship between stellar and planetary composition.
This is also the case for metrics such as density, or scaled density, as used in previous studies \citep{Adibekyan2021}.

Often when discussing planetary interiors, reference is made to four main layers: a predominantly iron core, a predominantly silicate mantle, a water layer, and an atmosphere.
For rocky planets, considering only the core and the mantle is a somewhat valid assumption, as thin secondary atmospheres do not make a significant difference to the planetary radius (though thicker atmospheres would affect this).
Therefore, using $M_\mathrm{p}\,R_\mathrm{p}^{-4}$ as a metric for the composition of super-Earths contains valid assumptions.
However, as shown in the clustering of sub-Neptunes in the bottom half of Fig.~\ref{fig:firon_mr4}, considering only the core and mantle layer is an invalid assumption for planets that have a significant envelope (water layer and an atmosphere).
These planets require a more intricate model to capture their more complex structures and overcome degeneracies before their compositions are compared to those of their host stars.
We discuss this in more detail in Section~\ref{sec:int}.

\section{Interior modelling}
\label{sec:int}

To compare the composition of a planet with that of its host star, we must quantify the planetary makeup. 
Specifically, we are interested in comparing the iron and silicate content, as these are the primary components of planetary cores and mantles.
Interior modelling codes allow us to make inferences about the mass fractions of elements within each planet.
In this work, we make use of \texttt{plaNETic}\footnote{\url{https://github.com/joannegger/plaNETic}} \citep{Egger2024}, an interior modelling framework that utilises neural networks, and is based on the \texttt{BICEPS} model \citep{Haldemann2024}.
With this package, we calculated various planetary layer mass fractions and estimated the amount of iron and silicates that may be present in the core and mantle.
We employ the model in which the planetary interior structure is computed independently of the host star chemical composition.
For each system, \texttt{plaNETic} requires the following parameters: the number of planets being modelled in the system; system age; stellar mass ($M_\star$); stellar radius ($R_\star$); stellar effective temperature (\teff); planetary orbital period ($P$); planetary mass ($M_\mathrm{p}$); and radius ratio ($R_\mathrm{p}/R_\star$).

We ran the water-rich \texttt{plaNETic} models for all planets in our sample, obtaining mass fractions for the core, mantle, water, and H/He contributions to the total mass, alongside the molar fractions ($x$) of specific elements within the core and the mantle.
The core is assumed to be comprised of iron (Fe) and sulphur (S), while the mantle contains Fe, magnesium (Mg), silicon (Si), and oxygen (O).
For more detail on the specific core and mantle compositions and the molecular forms that these elements take, refer to \citet{Haldemann2024} and references therein.
Using these outputs, the iron mass fraction of the entire planet (contributions from the core and the mantle) were calculated using the generalised equation
\begin{equation}\label{eq:fironp_total}
    f_\mathrm{iron}^\mathrm{planet} = \frac{M_\mathrm{Fe}x_\mathrm{Fe}}{\sum_i M_i x_i} \Bigg|_\mathrm{\,core} f_\mathrm{core} + \frac{M_\mathrm{Fe}x_\mathrm{Fe}}{\sum_j M_j x_j} \Bigg|_\mathrm{\,mantle} f_\mathrm{mantle}.
\end{equation}
Here, $M$ is the molar mass in the specified layer for a given element, $x$ is the molar fraction of a given element, $f_\mathrm{core}$ and $f_\mathrm{mantle}$ are the core and mantle mass fractions, respectively, $i$ denotes the core elements (Fe and S), and $j$ denotes the mantle elements (Fe, Mg, and Si, all sequestered in oxygen-bearing compounds).
This can be rescaled to only account for the mass contribution of the core and the mantle,
\begin{equation}\label{eq:fironp}
    \fironp = \frac{f^\mathrm{core}_\mathrm{Fe}+f^\mathrm{mantle}_\mathrm{Fe}}{f_\mathrm{core}+f_\mathrm{mantle}} = \frac{f_\mathrm{iron}^\mathrm{planet}}{f_\mathrm{core}+f_\mathrm{mantle}},
\end{equation}
which allows for comparison between the super-Earths and sub-Neptunes in the sample.
We then use this as a proxy for planetary composition.

\begin{figure}
	\includegraphics[width=0.99\columnwidth]{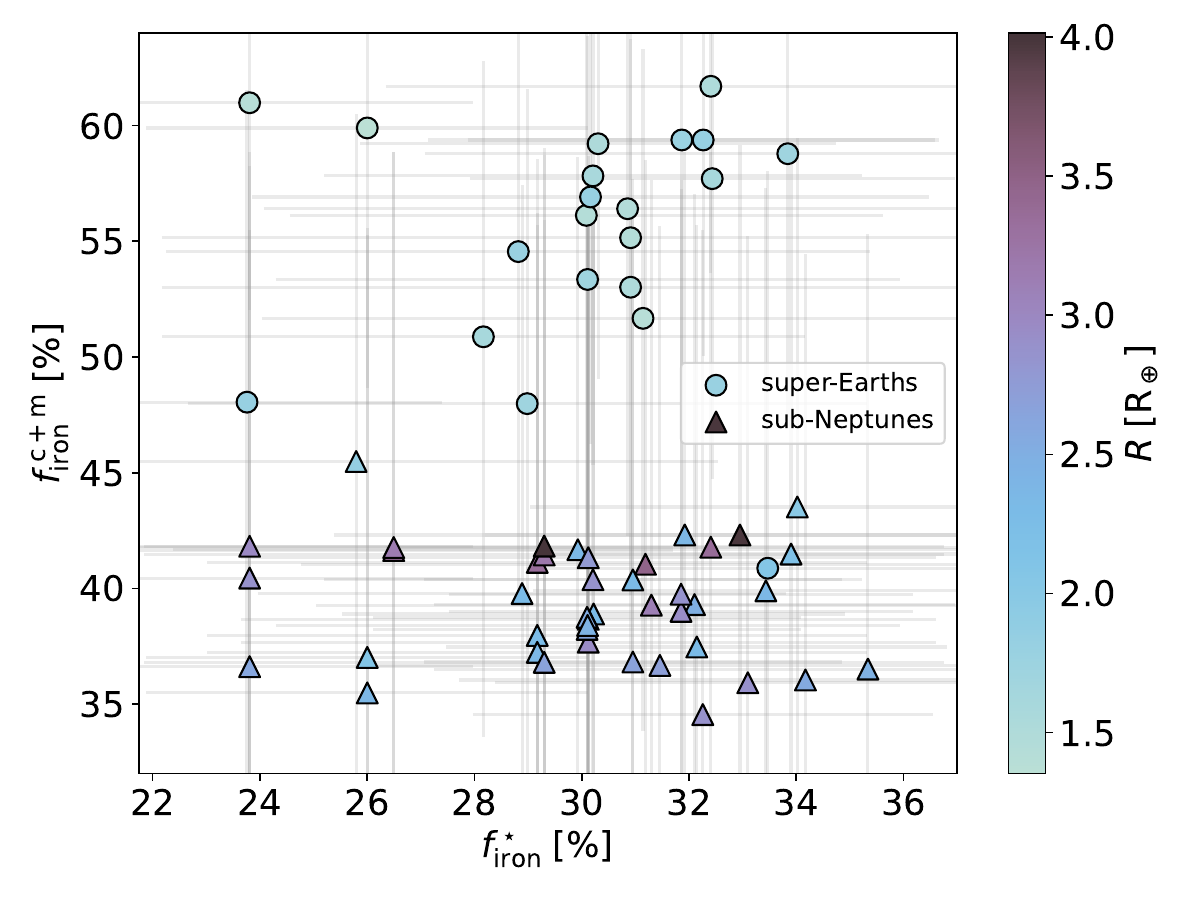}
    \caption{Planetary iron mass fraction vs. stellar iron mass fraction. Super-Earths and sub-Neptunes are shown as circles and triangles, respectively.}
    \label{fig:st_pl_comp}
\end{figure}
The results of this analysis are shown in Fig.~\ref{fig:st_pl_comp}, where \fironp is the fraction of the planetary mass that is due to iron, considering the abundance both in the core and the mantle.
As expected, Super-Earths occupy the high-\fironp region, meaning a larger proportion of their mass is attributed to iron.
Sub-Neptunes, with significant envelopes, occupy the low-\fironp region.
Following previous studies, we fit a linear function ($y=mx+c$) to the super-Earth population, with $y=\fironp$ and $x=\firons$, using the \texttt{ODR} routine from \texttt{scipy}. We separately fit a linear function for the sub-Neptune population as they are intrinsically different from super-Earths.
\begin{table}
        \centering
        \caption{Linear fit results for the super-Earth and sub-Neptune populations.} 
        \renewcommand{\arraystretch}{1.4}
        \begin{tabular}{lcccc}
        \hline
        \hline
             Population & slope ($m$) & intercept ($c$) & $r^2$ & $\chi^2$ \\ 
             \hline
             Super-Earths & $-0.006 \pm 0.317$ & $57.613 \pm 9.588$ & 0.0005 & 0.161 \\
             Sub-Neptunes & $-0.113 \pm 0.133$ & $42.887 \pm 4.009$ & 0.0141 & 0.018 \\
             \hline
        \end{tabular}
        \label{tab:ymxc}
\end{table}

Neither population exhibits a correlation, with slopes consistent with zero, and Pearson's $r^2$ values close to zero ($r^2=0.0005$ and $r^2=0.0141$ for super-Earths and sub-Neptunes, respectively). The individual fit statistics are listed in Table~\ref{tab:ymxc}. This significant lack of a correlation between stellar and planetary iron mass fraction is in stark contrast to \citet{Adibekyan2021, Adibekyan2024}, but consistent with the most recent result of \citet{Ross2025}.

The planetary distribution separated by the radius valley in Fig.~\ref{fig:rad_valley} is highly consistent with the bimodality shown in Fig.~\ref{fig:st_pl_comp}, with super-Earths lying in the high-\fironp region and sub-Neptunes in the low-\fironp region.
Additionally, the colour gradient of Fig.~\ref{fig:rad_valley} highlights the sharp transition in \fironp between super-Earths and sub-Neptunes, as opposed to a smooth gradient.
This echoes the sharp changes in bulk density which \citet{Luque2022} find between sub-Neptunes, water worlds and super-Earths orbiting M dwarfs. We also do not find evidence of intermediate-density water worlds.
However, the sample shown here is different as we use a more complex parameter than the bulk density and focus on systems with FGK host stars.
The presence of these two distinct populations in our study suggests that sub-Neptunes have a more complex interior structure than a super-Earth-like core with an added envelope.

\subsection{Specific cases}

While PlanetS contains the recommended papers for each planet/system, a large number of the planets that are described above have contested masses, with some planets having published mass ranges of up to $\sim9\,\mathrm{M_\oplus}$\footnote{using masses from the NASA Exoplanet Archive (\url{https://exoplanetarchive.ipac.caltech.edu})}.
A change in planet mass of $1\,\mathrm{M_\oplus}$ can lead to a deviation in the calculated \fironp of around 5 per cent.
Some parameter combinations, when inputted and ran via \texttt{plaNETic}, result in a model run discarding >99.9 per cent of the generated instances, leaving less than 10\,000 samples in the outputted posterior distributions.
This is likely due to \texttt{plaNETic} assuming that each planet hosts at least a minimum amount of H and He, with a mass fraction of $10^{-6}$ relative to the total planet mass.
While such a small mass fraction is physically motivated (layers smaller than this would be unstable against evaporation), it becomes problematic for particularly dense planets where even this minimal envelope pushes the predicted radius above the observed value.
Alternatively, measured planetary masses could be over-estimated, meaning that the accuracy of the parameters themselves could be hindering the analysis.
Below, we discuss the existing literature surrounding these planets and examine to what extent running \texttt{plaNETic} with different parameter combinations can affect the outputs.

\subsubsection{Dense planets}

\begin{figure}
	\includegraphics[width=0.95\columnwidth]{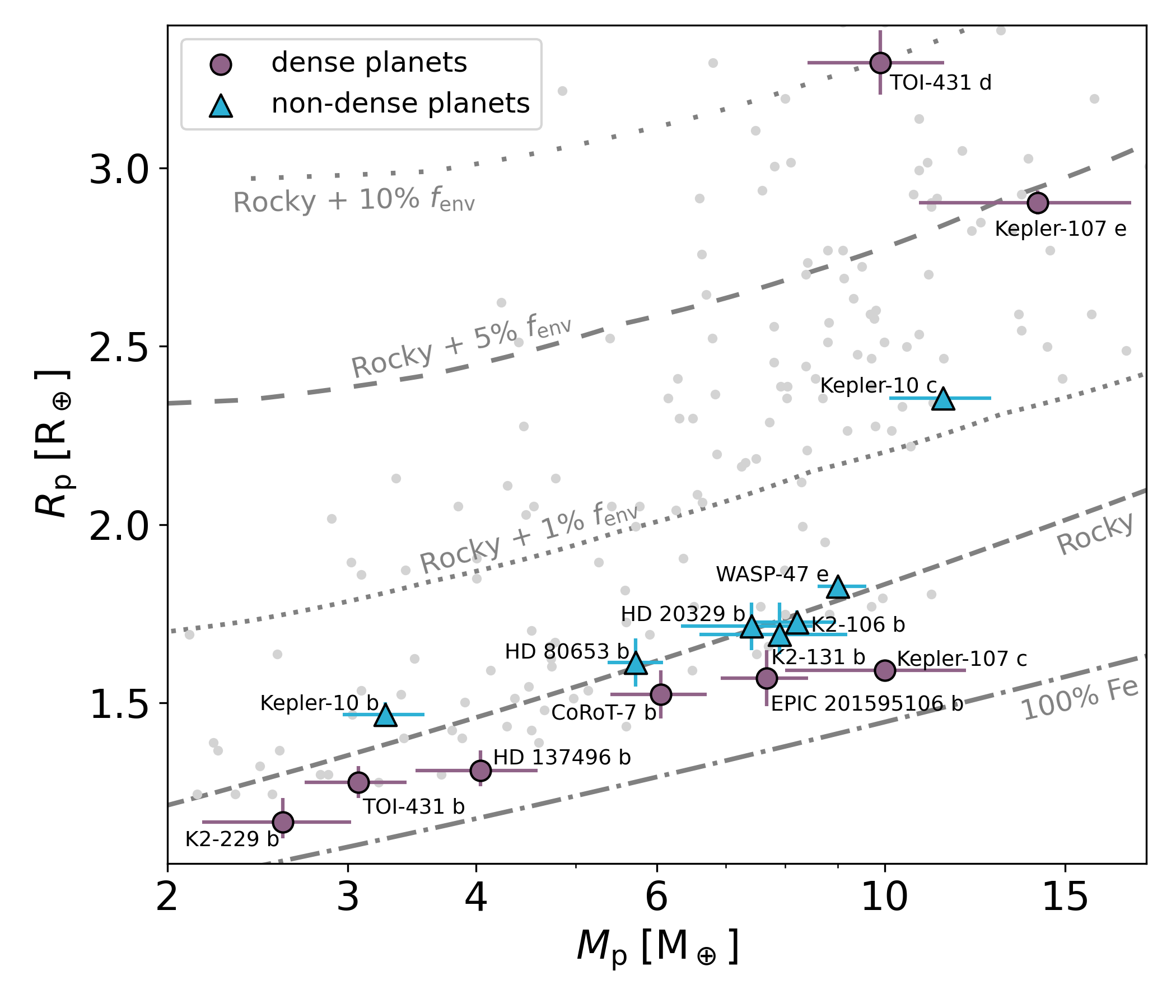}
    \caption{Mass--radius diagram highlighting the systems that were re-visited after analysis via \texttt{plaNETic} (background points indicate the remainder of the sample). Purple circles represent systems where one of the planets was identified to be particularly dense using the set of parameters reported in PlanetS. Blue triangles are systems that may have failed for other reasons. As in Fig~\ref{fig:mr}, compositional lines are from \citet{Zeng2019} for the 100\% iron and rocky, Earth-like lines, and \citet{Chen2016} for the rocky lines with various envelope mass fractions.}
    \label{fig:mr_dense}
\end{figure}

Fig.~\ref{fig:mr_dense} shows planets where >99.9 per cent of samples were discarded for their particular \texttt{plaNETic} runs.
Around half of these planets reside in systems where one of the planets has a relatively high density (shown as purple circles).
Since this region of parameter space is sparsely sampled, \texttt{plaNETic} is not able to model denser planets as effectively as those with Earth-like or Neptune-like densities.
In the case of TOI-431\,d, since multi-planetary systems are modelled in \texttt{plaNETic} as a system, TOI-431\,b caused its removal as it is too dense.

\textbf{CoRoT-7\,b:} 
At the time of writing, the most precise mass measurement for CoRoT-7\,b is $M_\mathrm{p}=6.056\pm0.653\,\mathrm{M_\oplus}$, obtained by \citet{John2022}, hence why this set of parameters is included in the PlanetS database.
Using the planetary parameters presented in \citet{John2022}, \texttt{plaNETic} retains only 4\,203 instances out of 10$^7$, compared to 146\,696 out of 10$^7$ when using those from \citet{Stassun2017}.
The planetary parameters from \citet{Stassun2017} were subsequently adopted as the default; this alternative mass and radius result in a smaller density than in \citet{John2022}.

\textbf{K2-229\,b:}
\texttt{plaNETic} was run using the planetary parameters from \citet{Santerne2018}, which resulted in 410 samples remaining for the water-rich model and 343 for the water-poor model.
Running \texttt{plaNETic} with an alternative set of planetary parameters would not have much effect as, in the literature, the two other masses that exist for K2-229\,b \citep{Dai2019, Howard2025} are consistent with that of \citet{Santerne2018}.
As a result, K2-229\,b was removed from the sample.

\textbf{Kepler-107\,c and e:}
For this system, neither model A nor model B for either planet (c or e) retain enough instances to be included in further analysis.
As discussed in \citet{Bonomo2019}, Kepler-107\,c has a density above average of what is expected for a planet of its size.
\citet{Bonomo2019} hypothesised that this density is the product of a giant impact event resulting in the mantle being stripped off the planet, leaving behind a denser iron core.
As such, Kepler-107\,c may be a bare, atmosphere-free core, which falls outside the scope of \texttt{plaNETic}'s model grid.
Subsequently, we modelled Kepler-107\,c and e as separate systems; these runs yielded 337\,503 samples for planet e but still only $\sim$7000 for planet c, as expected.
Kepler-107\,c was removed from the sample as a result.

\subsubsection{Single-paper systems}

In the case of EPIC 201595106\,b \citep{Livingston2024}, HD 20329\,b \citep{Murgas2022}, HD 137496\,b \citep{AzevedoSilva2022}, and TOI-431\,b and d \citep{Osborn2021}, only one analysis per system has been completed to obtain the masses of the planets.
All of these systems have less than 10\,000 samples remaining at the end of the runs.
As there are no alternative parameter sets with which to run \texttt{plaNETic}, we exclude these systems from the study.

\subsubsection{HD 80653\,b}

HD 80653\,b is a super-Earth, with its mass characterised first by \citet{Frustagli2020} and then followed by \citet{Bonomo2023}.
The \texttt{plaNETic} run using the \citet{Bonomo2023} planetary parameters retained only 181 instances for the water-poor model, but over 70\,000 for the water-rich model.
Planets modelled with the water-rich prior are likely to be water-dominated and therefore have a higher mean molecular mass atmosphere.
This makes it easier to gravitationally retain an atmosphere, hence why the water-rich prior performed better than the water-poor prior.

The two analyses used the same RVs from HARPS-N and resulted in masses that are consistent to $1\sigma$.
A \texttt{plaNETic} analysis was therefore not repeated for the \citet{Frustagli2020} mass, since this would have resulted in similar outputs, and it was subsequently removed from this study.

\subsubsection{K2-106\,b}

The mass included in PlanetS is that from \citet{Bonomo2023}, which results in only 48 instances remaining from the \texttt{plaNETic} analysis.
When re-running the analysis with planetary parameters from \citet{Palethorpe2026}, even fewer samples were retained (24).
This planet was removed from any following analysis.

\subsubsection{K2-131\,b}

\citet{Bonomo2023} found that the mass of K2-131\,b is $M_\mathrm{p}=7.9\pm1.3\,\mathrm{M_\oplus}$ and the radius is $R_\mathrm{p}=1.690^{+0.085}_{-0.058}\,\mathrm{R_\oplus}$.
Running \texttt{plaNETic} with this planetary parameter configuration yields posteriors based on only 7334 samples.
The number of samples can be increased to 16\,920 by using the planetary parameters from \citet{Dai2019} ($M_\mathrm{p}=6.3\pm1.4\,\mathrm{M_\oplus}$; $R_\mathrm{p}=1.651^{+0.065}_{-0.056}\,\mathrm{R_\oplus}$).
These planetary parameters are adopted as default for the rest of the study.

\subsubsection{Kepler-10\,b and c}

Kepler-10 is a well-studied system, with five papers reporting masses for both planet b and c; the range of masses reported for Kepler-10\,b spans $1.37\,\mathrm{M_\oplus}$.
The masses from \citet{Bonomo2023} were selected for inclusion in the PlanetS catalogue, and were hence used as inputs for a \texttt{plaNETic} run, retaining only 1601 samples for the water-poor model.
Re-running with a new set of parameters \citep{Weiss2016}, the water-poor model keeps only 703 instances and was subsequently removed from the sample.

\subsubsection{WASP-47\,e}

For WASP-47\,e, the characterisation included in the PlanetS database is that by \citet{Nascimbeni2023}.
In comparison to a number of other mass characterisation papers on this planet, \citet{Nascimbeni2023} report a relatively large mass of $M_\mathrm{p} = 9.0^{+0.6}_{-0.4}\,\mathrm{M_\oplus}$.
An initial \texttt{plaNETic} run with these parameter values yielded posteriors with only 2590 samples. Re-running with the masses and radii given by \citet{Bryant2022} results in posteriors of length 22\,845.

\section{Discussion}
\label{sec:disc}

\subsection{Parameter precision}

A persistent issue that hinders the comparison of stellar and planetary composition is the systematic underestimation of uncertainties and their improper propagation \citep{Freckelton2026}.
This leaks through into the errors of the metrics that are used to quantify stellar composition, i.e., \firons, meaning these values are also underestimated.
This problem is further exacerbated in the case of relative abundance ratios such as \alphafe.
For these ratios, a complete error analysis must not only account for the atomic and photospheric errors, but also for the uncertainties inherent in the solar abundances used to convert from absolute to relative stellar abundances.
On an element-by-element basis, accounting for the solar errors would result in a constant shift in values, but when the solar uncertainties are propagated through to compositional proxies (e.g., [Si/Fe], [Mg/Fe], \alphafe), this is not the case.
Omitting this solar contribution means a source of error is unaccounted for, hence the precision of stellar abundances quoted in literature is often incorrect.

Conclusions drawn from the data displayed in Figs.~\ref{fig:firon_mr4} and \ref{fig:st_pl_comp} are impacted by the precision to which we know the values themselves.
As these stars host precisely characterised planets, we have abundant, high-S/N spectra; the quality of the observational data is not the limiting factor.
Rather, the precision of compositional studies is fundamentally limited by the systematic uncertainties inherent to the stellar atmospheric models used in spectral analysis.
Even with thorough error propagation, these systematics dominate errors, placing a ceiling on the conclusions that can be drawn, regardless of data quality.
With average absolute errors on \firons of $\sim$5 per cent, the uncertainties quoted in this paper are larger than those in previous studies.
However, these errors incorporate the true, inherent uncertainties on the stellar abundances and their related proxies.

\subsection{Parameter accuracy}

A key shortcoming in many previous studies is the treatment of planetary parameter values as true and absolute, despite inconsistencies between different system analyses being a known issue.
This is particularly prevalent when looking at the consistency of planetary masses when using differing observational datasets and methods \citep{Osborne2025}. 
The same concern applies not only to masses but also to radii; examples include TOI-1685 b \citep{Bluhm2021, Egger2025, Luque2025} or TOI-238 b \citep{Mistry2024, SuarezMascareno2024, Egger2025}.
Subsequently, without taking this into consideration, studies are built on a sample of systems for which the multiple sets of planetary parameters often conflict.
When planetary parameters such as mass change, this can have large effects on the conclusions that we draw about their compositions.

The cases of Kepler-10\,b and TOI-561\,b are pertinent examples of how planetary mass can vary widely between publications.
First measured by \citet{Batalha2011}, the mass of Kepler-10\,b currently has a reported mass range of $1.37\,\mathrm{M_\oplus}$ \citep{Esteves2015, Bonomo2025}.
For TOI-561\,b, the planetary masses reported in two of the most referenced papers, \citet{Lacedelli2022} and \citet{Polanski2024}, are $M_\mathrm{p} = 2.00\,\mathrm{M_\oplus}$ and $M_\mathrm{p} = 3.00\,\mathrm{M_\oplus}$, respectively.
With a relative difference of 42 and 50 per cent for Kepler-10\,b and TOI-561\,b, respectively, these discrepancies can drastically change the inferred composition of the planets.
This is particularly consequential for TOI-561\,b as it is frequently cited in studies exploring links between planetary and stellar composition, owing to its host star being metal-poor and alpha-enhanced \citep[e.g. ][]{Adibekyan2021,Adibekyan2024}.
The recent inference of a secondary atmosphere on TOI-561\,b by \citet{Teske2025} also has broader implications.
If this result is representative of the super-Earth population as a whole, it would suggest that volatile content, which exerts a substantially greater influence on mass-radius relations and CMF estimates than rocky interior composition, may systematically overprint any underlying correlation between planetary and stellar abundances.
Atmospheres of varying thickness would mask bulk density trends driven by stellar composition alone, providing a physically motivated explanation for the absence of a clear observational relationship between CMF and host star abundances.
This compositional degeneracy warrants careful scrutiny of the conclusions drawn from such analyses.

While we reduce the extent of the inhomogeneities present in the planetary parameters used in this study by re-fitting for new stellar parameters (see Sections~\ref{sec:stelparam} and \ref{sec:link}), they cannot be removed completely.
Planetary fitting methods are inconsistent between studies due to the many required methodological decisions: choice of fitting code, assumptions regarding the number of planets in a system, and whether (and how) GPs are employed to mitigate stellar activity.

It is also important to consider the accuracy of the stellar abundances employed in this analysis. We find no spurious trends between \teff, $\log{g}$, or $\xi$ and \feh, \mgh, and \sih, indicating that no systematic biases are identifiable in our sample. Additionally, since the method used to derive stellar abundances in this work assumes LTE, it is also necessary to evaluate the extent to which departures from this assumption might affect the conclusions drawn. For dwarfs with near-solar metallicities, departures from LTE rarely have a significant effect on the derived parameters or abundances for the lines considered here. This has been discussed in literature, both for the iron lines used and also for other species such as magnesium and silicon \citep[e.g.,][]{Lind2012, Ruchti2013, Tsantaki2013, Adibekyan2015, Mikolaitis2019, Lind2024}. In particular, \citet{Lind2012} show that non-LTE (NLTE) corrections for Fe\,\textsc{i} and Fe\,\textsc{ii} lines in FGK dwarfs remain below $\sim$0.05 dex for $\feh\gtrsim -1$, well within our typical abundance uncertainties, and that the corresponding effect on \teff and $\log{g}$ derived from excitation/ionisation balance is similarly small and within our errors. Given that our sample is restricted to near-solar metallicity dwarfs, we therefore expect NLTE effects to be negligible compared to our quoted uncertainties.

\subsection{Benchmarking to the Solar System}

To test the reliability of the \fironp values calculated from the \texttt{plaNETic} outputs, we ran \texttt{plaNETic} on Earth and Uranus; this subset comprises Solar System planets whose masses fall within the range that \texttt{plaNETic} can model ($0.5\,\mathrm{M_\oplus}<M_\mathrm{p}<15\,\mathrm{M_\oplus}$) and that cover the approximate radius range studied in this work ($1 \,\mathrm{R_\oplus} \lesssim R_\mathrm{p} \lesssim 4\,\mathrm{R_\oplus}$).
Comparing the output values of, for example, iron mass fraction with the known values for the Solar System planets allows us to contextualise the result of the wider sample.

From literature, the mass fraction of iron within the Earth is approximately 32 per cent; 85.0 per cent of the core ($1.93\times10^{24}\,\mathrm{kg}$) by mass is iron and 6.2 per cent of the mantle ($4\times10^{24}\,\mathrm{kg}$) by mass is iron \citep{DeHoog2010, McDonough2017}.
From \texttt{plaNETic}, we obtain a value of $47^{+9}_{-13}$ per cent for the water-rich case and $56^{+9}_{-13}$ per cent for the water-poor case, for the amount of Earth that is comprised of iron by mass (using Equation~\ref{eq:fironp}).
As aforementioned, though these values are larger than the literature ranges for Earth, they provide a benchmark for the other planets in the sample.
It is likely that these values are large due to the amount of iron present in the modelled mantles.

While we know the interior of Uranus better than we know the interiors of exoplanets, we do not have a measurement of its iron mass fraction.
Regardless, the resulting masses of iron in Uranus, obtained from \texttt{plaNETic}, are $25^{+14}_{-12}$ and $36^{+14}_{-17}$ per cent (water-rich and water-poor models, respectively).
The corresponding \fironp values are $42^{+17}_{-20}$ per cent for both model types.
For the purposes of this study, Earth is categorised as a rocky planet/super-Earth and Uranus is categorised as a gaseous planet/sub-Neptune.

In comparison to the remainder of the sample, the \texttt{plaNETic} \fironp values for Earth are lower than the super-Earth sample average, with the average \fironp value for super-Earths being $56^{+9}_{-14}$ per cent and $60^{+7}_{-10}$ per cent for the water-poor and water rich models, respectively.
It is important to note that this trend may just be an effect of the assumptions underlying \texttt{BICEPS}/\texttt{plaNETic}, rather than a true physical signal.
However, if this is a physical signal, we speculate that this may be due to \fironp having some dependence on the period of rocky planets, or more likely, their equilibrium temperatures.
One possibility is that iron becomes more abundant relative to silicates interior to the silicate sublimation line, though whether rocky planets can form at such close separations remains uncertain \citep{Chiang2013, Chatterjee2014, Owen2017}.
Alternatively, shorter-period planets may experience more frequent collisions, which can strip silicate mantles and enrich the iron fraction.
Therefore, planets with shorter periods may result in larger planetary iron mass fractions than those that reside on longer orbits.
This potential link with equilibrium temperature is explored in the following section.
In the case of sub-Neptunes, the average \fironp values are $39^{+18}_{-19}$ per cent for both models, meaning Uranus is largely representative of the sub-population.

\subsection{Relation with equilibrium temperature}

\texttt{plaNETic} does not account for equilibrium temperature directly, but orbital period is given as an input parameter \citep{Egger2024}, and will indirectly encompass the effects of equilibrium temperature.
As such, to test any possible dependence on equilibrium temperature, the Earth was run with \texttt{plaNETic} for two alternative period values: $P=3$ days and $P=10$ days.
We found that the closer in the planet is to its star, the higher its \fironp (for the water-rich priors), with the results for $P=3$ days yielding \fironp values of $66^{+4}_{-6}$ per cent and $42^{+24}_{-21}$ per cent, and the results for $P=10$ days yielding values of $65^{+5}_{-7}$ per cent and $45^{+18}_{-25}$ per cent, for the water-rich and water-poor priors, respectively.
For water-poor priors, the resultant values decrease with shorter orbital periods; however, the associated uncertainties continue to encompass the values derived for Earth at a period of $P=365$ days, meaning the differences are statistically insignificant.

The way in which a planetesimal accretes material in a protoplanetary disc is influenced by the local disc temperature at its formation site; planetesimals forming in warmer, inner disc regions may, on average, accrete differently to those forming further out.
It is important to note, however, that disc conditions evolve over time, planets can migrate from their formation locations, and disc temperature does not equate to planetary equilibrium temperature.
Nevertheless, close-in planets may generally also form further in.
As such, equilibrium temperature remains a useful observational proxy for the broader thermal environment a planet inhabits, and investigating the relationship between \firons, \fironp, and equilibrium temperature may still provide insight into the link between stellar and planetary composition.

\begin{figure}
	\includegraphics[width=0.99\columnwidth, trim={2cm 0 1.5cm 0}, clip]{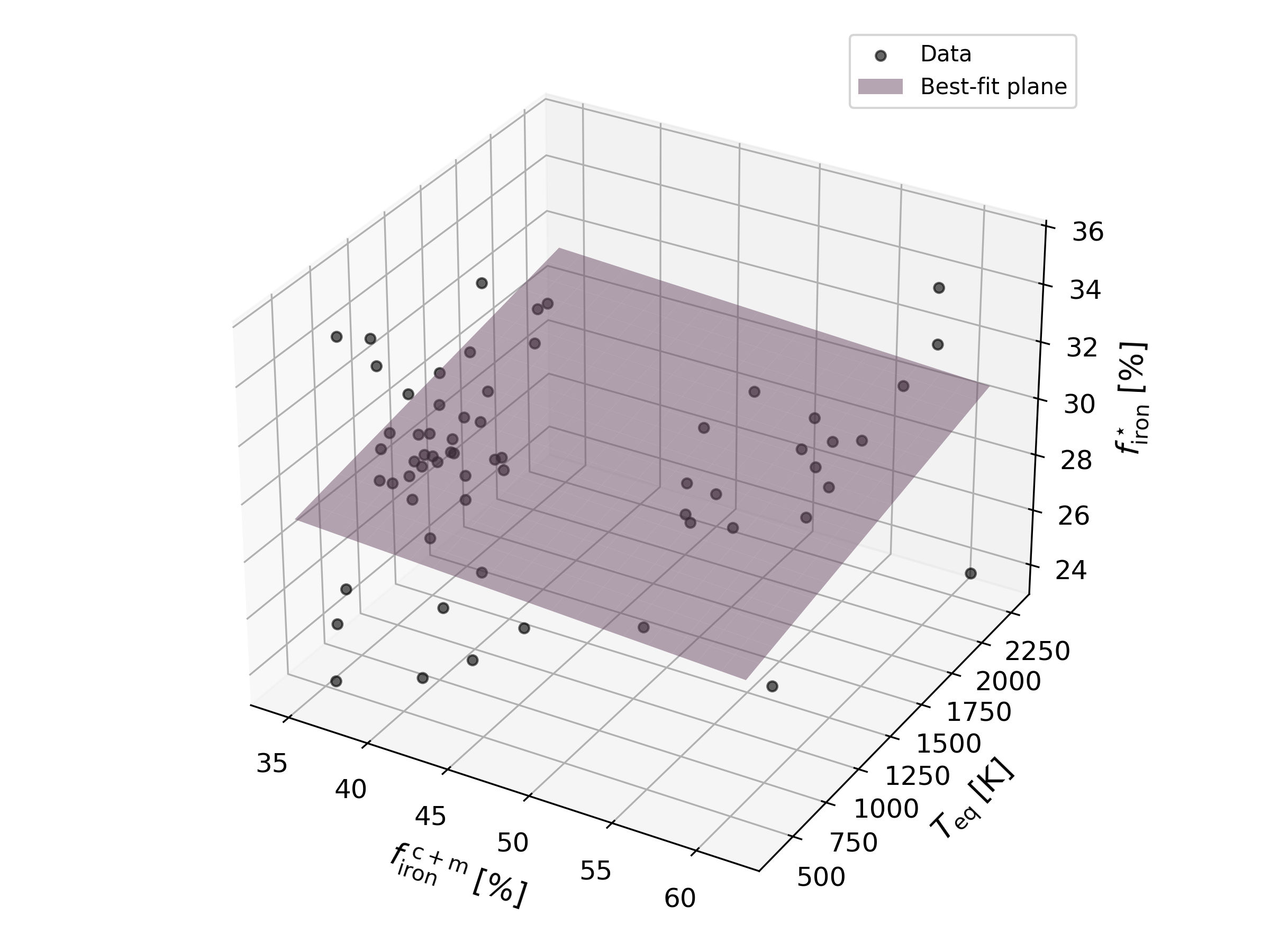}
    \caption{Plane fit to the entire sample using the alpha parameterisation (Equation~\ref{eq:alpha}).}
    \label{fig:all_alpha}
\end{figure}

\begin{figure}
	\includegraphics[width=0.99\columnwidth, trim={2cm 0 1.5cm 0}, clip]{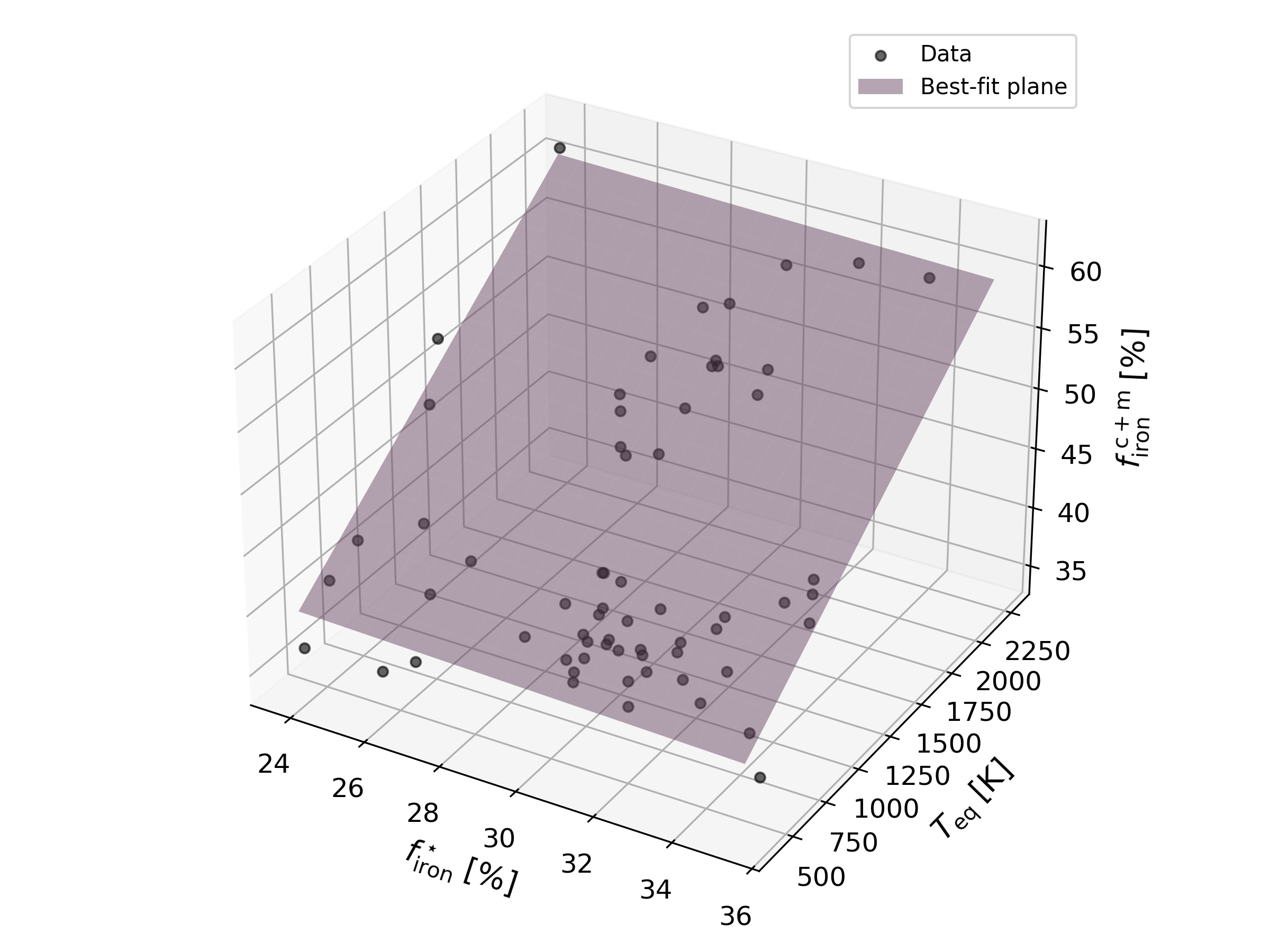}
    \caption{Plane fit to the entire sample using the beta parameterisation (Equation~\ref{eq:beta}).}
    \label{fig:all_beta}
\end{figure}

\begin{figure}
	\includegraphics[width=0.99\columnwidth, trim={2cm 0 1.5cm 0}, clip]{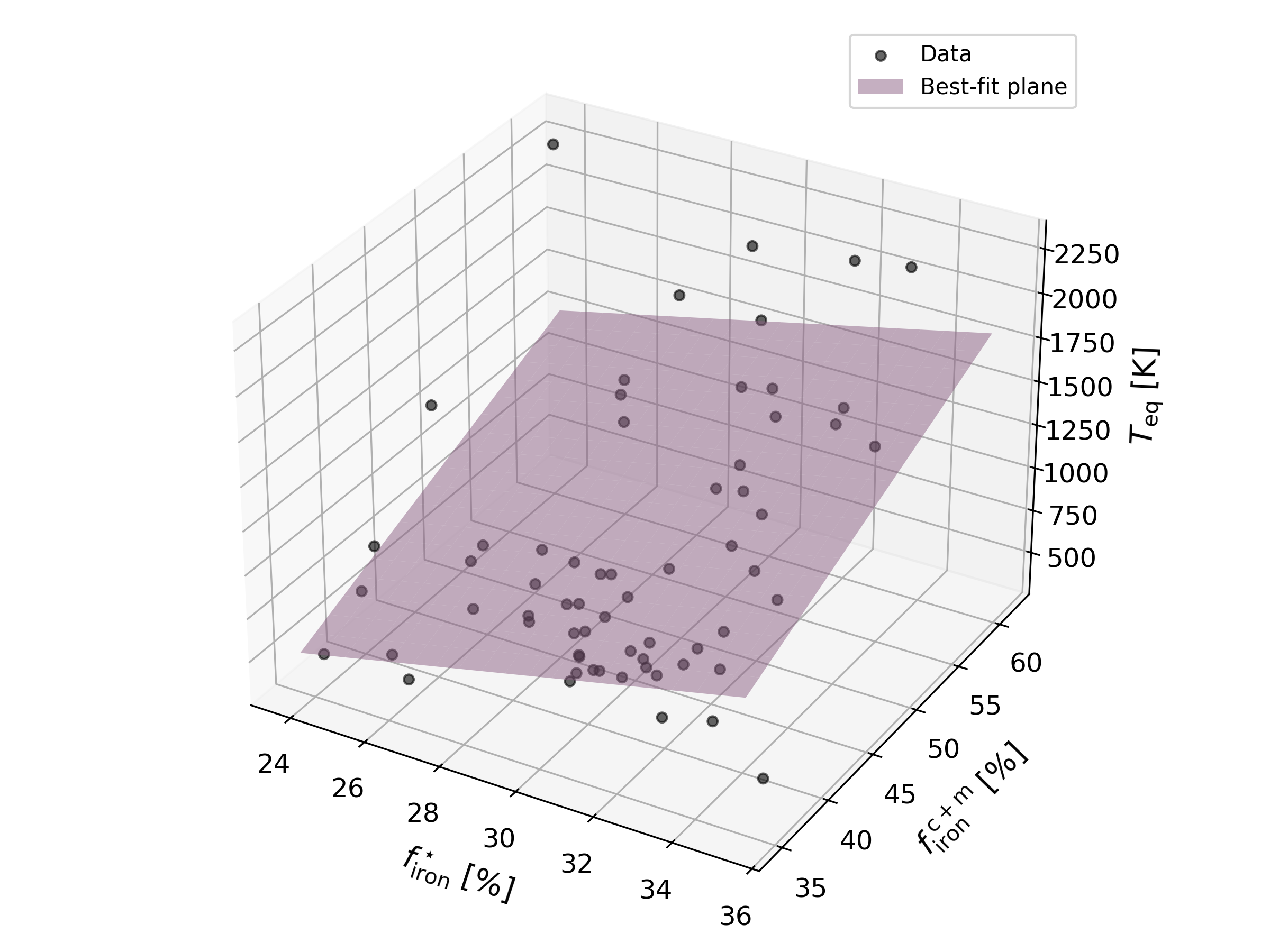}
    \caption{Plane fit to the entire sample using the gamma parameterisation (Equation~\ref{eq:gamma}).}
    \label{fig:all_gamma}
\end{figure}

Using \texttt{emcee}, we fit planes with three parameterisations:
one in \firons--$T_\mathrm{eq}$--\fironp space (alpha parameterisation),
\begin{equation}\label{eq:alpha}
    \firons = \alpha_0 + \alpha_1T_\mathrm{eq} + \alpha_2\fironp;
\end{equation}
one in \fironp--\firons--$T_\mathrm{eq}$ space (beta parameterisation),
\begin{equation}\label{eq:beta}
    \fironp = \beta_0 + \beta_1\firons + \beta_2T_\mathrm{eq};
\end{equation}
and the final one in $T_\mathrm{eq}$--\firons--\fironp space (gamma parameterisation), 
\begin{equation}\label{eq:gamma}
    T_\mathrm{eq} = \gamma_0 + \gamma_1\firons + \gamma_2\fironp.
\end{equation}
Here, $\alpha_0$, $\beta_0$, $\gamma_0$, and  are intercepts, and $\alpha_1$, $\alpha_2$, $\beta_1$, $\beta_2$, $\gamma_1$, and $\gamma_2$ are gradients.
The values of these parameters are shown in Tables~\ref{tab:planes} and \ref{tab:sub_planes}, and plane fits for the entire sample are shown in Figs.~\ref{fig:all_alpha}, \ref{fig:all_beta}, and \ref{fig:all_gamma}.

For both the alpha and beta parameterisations, planetary composition and stellar composition are able to be directly compared ($\alpha_2$ and $\beta_1$).
Considering the values for the entire population, both are consistent with zero; no evidence of any connection exists.
This is also apparent when fitting a plane to the two sub-populations (super-Earths and sub-Neptunes; see Appendix~\ref{sec:plane}).

\begin{figure*}
	\includegraphics[width=0.99\textwidth]{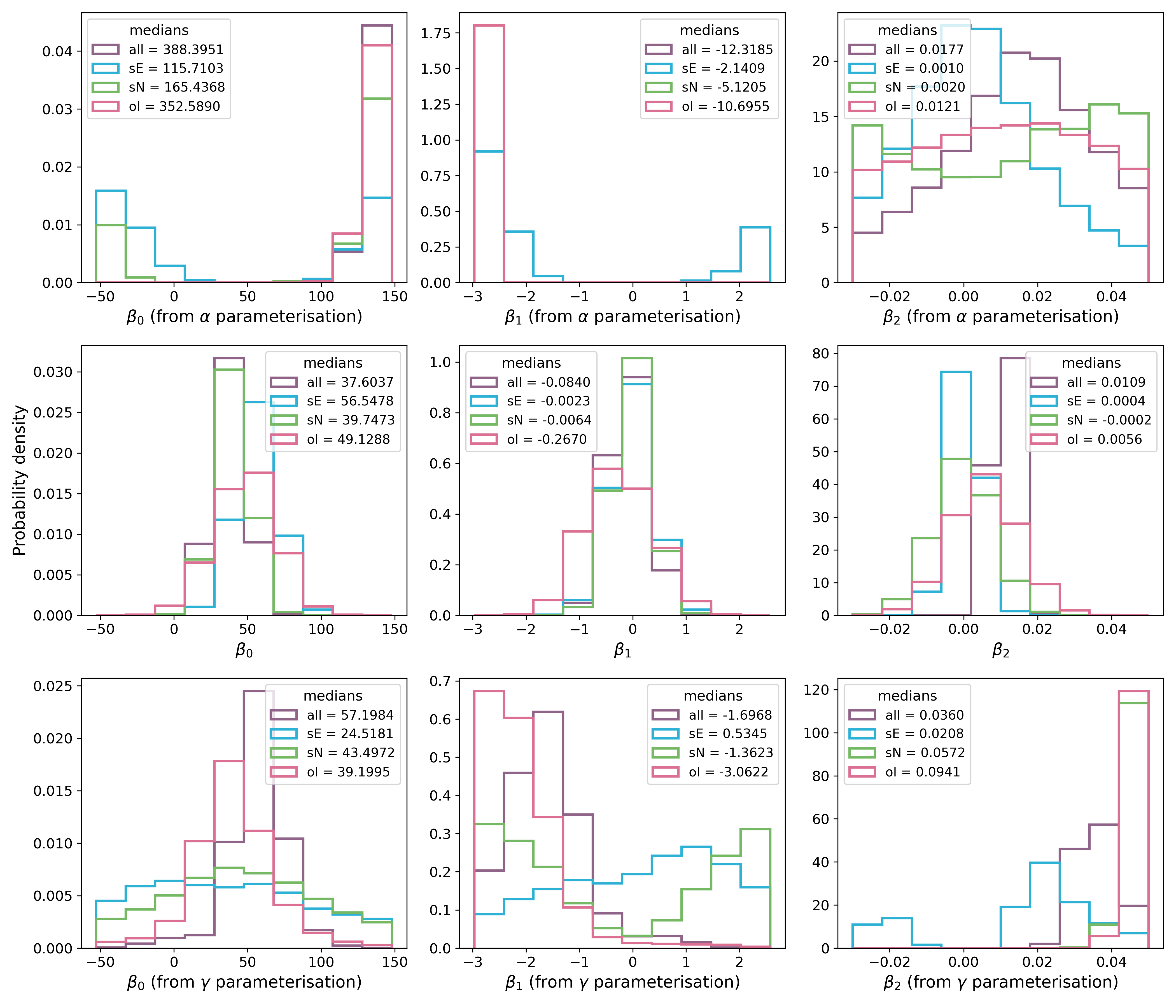}
    \caption{Comparison between the $\beta$ values calculated from the three parameterisations (top row: $\alpha$; middle row: original $\beta$; bottom row: $\gamma$).}
    \label{fig:beta_hists}
\end{figure*}

For a relationship between \firons, \fironp, and $T_\mathrm{eq}$ to be considered robust, fitting a reparameterised plane and subsequently rearranging the derived parameters to recover one parameterisation from another must yield consistent values.
Table~\ref{tab:re} presents the values for the beta parameterisation, derived from the values of the coefficients from the alpha and gamma parameterisations.
Summaries of all of the distributions are shown in Figure~\ref{fig:beta_hists}.
All three parameterisations result in dissimilar values for the beta coefficients.
The alpha parameterisation in particular yields results that are unconstrained, with all instances ($\beta_0$, $\beta_1$, and $\beta_2$) being consistent with zero.
Using the gamma parametrisation delivers more constrained coefficients, however none of the values are consistent with the original values from the beta parameterisation.
This is indicative of the fact that the relations drawn from this sample are not robust, and are instead sensitive to the choice of parameterisation.
As such, any physical interpretation of the results from the three-dimensional fits should be treated with caution, as the inferred relations appear to be driven by the parameterisation adopted, rather than by the underlying data.

Nevertheless, there are tentative indications that \fironp and \firons retain some dependence on $T_\mathrm{eq}$.
Equation~\ref{eq:gamma} links $T_\mathrm{eq}$ with \fironp and \firons.
All of the corresponding coefficients for this parameterisation are positive, non-zero and significant to $3\sigma$, suggesting that equilibrium temperature may still play a role in shaping the iron content of these planets.

The gamma parameterisation (Equation~\ref{eq:gamma}) indicates that both \firons and \fironp increase with $T_\mathrm{eq}$: the hottest planets tend to be the most iron-rich, and to orbit the more iron-rich stars.
The planetary side of this trend runs in the direction expected from thermally driven mass loss, whereby close-in, highly irradiated planets preferentially shed their lightest material and so present a higher iron mass fraction.
For the sub-Neptunes, this is most plausibly loss of the H/He envelope through photoevaporation or core-powered mass loss \citep{Owen2013, Gupta2019}, which reduces the radius and thereby raises the \fironp inferred by \texttt{plaNETic} rather than reflecting a genuine change in the rock-to-iron ratio; outright vaporisation and escape of silicate material, which would truly fractionate iron from silicates, requires far more extreme conditions and is expected to operate only at the lowest masses and highest temperatures \citep{Curry2024}.
Conversely, even among the nominally rocky super-Earths, the retention or outgassing of a secondary volatile envelope, as recently inferred for the ultra-hot super-Earth TOI-561\,b \citep{Teske2025}, can influence the mass--radius relation and the inferred CMF more strongly than the rocky composition itself.
Secondary envelopes of varying thickness would therefore mask any bulk-density trend set by stellar composition alone, offering a physically motivated explanation for the absence of a clear relationship between \fironp and \firons.
The planetary iron mass fraction is in any case an imperfect tracer of inherited composition, since internal redox evolution and magma-ocean circulation can entrain iron and modify the apparent core mass fraction independently of the host star \citep{Lichtenberg2021}.
Disentangling these competing effects from a primordial compositional signal is difficult, since the same irradiation also correlates with the super-Earth/sub-Neptune divide and with the layered assumptions of the interior model itself \citep{Young2025, Lichtenberg2025}.
We therefore caution that, while the \emph{direction} of the $T_\mathrm{eq}$ correlation is suggestive, the current precision on the compositional proxies does not permit these contributions to be cleanly separated.

However, with the current precision on the compositional proxies, this effect cannot be reliably constrained across all parameterisations.
It should be noted that the other coefficients linking these quantities (namely, $\alpha_1$ and $\beta_2$) are expected to be small, since \fironp and \firons are percentages whilst $T_\mathrm{eq}$ spans a range of $\sim$500--$2250\,\mathrm{K}$.
This is therefore a trend worth revisiting as sample sizes grow through dedicated observations efforts \citep[e.g., ][]{Eschen2025, Palethorpe2026}, or if more precise compositional proxies become available.

To ensure that this was not a sampling effect caused by the paucity of short-period sub-Neptunes and long-period super-Earths, we created a sub-sample, hereafter referred to as the overlapping region.
This is defined as the planets located in the $T_\mathrm{eq}$ region between the hottest sub-Neptune and the coolest super-Earth.
We see that the overall conclusions drawn from the fits to the entire population still hold with this smaller sample, suggesting that the overall trends observed are not a product of the asymmetric distribution of sub-Neptunes and super-Earths in $T_\mathrm{eq}$-space, but reflect a genuine feature of the population.
The individual plane fits for the overlap region, super-Earths, and sub-Neptunes are included in Appendix~\ref{sec:plane}.

\begin{table}
        \centering
        \caption{Best fit values for the three plane parameterisations (entire sample and overlapping region).}
        \renewcommand{\arraystretch}{1.4}
        \begin{tabular}{lcc}
        \hline
        \hline
             Parameters & Entire population & Overlapping region \\ 
             \hline
             $\alpha_0$ & $29.80^{+2.57}_{-2.49}$ & $31.17^{+4.64}_{-4.65}$ \\
             $\alpha_1$ & $(12^{+15}_{-16})\times 10^{-3}$ & $(0.8\pm3.8)\times10^{-3}$ \\
             $\alpha_2$ & $-0.03\pm0.06$ & $-0.03\pm0.07$ \\
             \hline
             $\beta_0$ & $37.60^{+10.91}_{-11.10}$ & $49.13^{+19.83}_{-21.41}$ \\
             $\beta_1$ & $-0.08^{+0.35}_{-0.35}$ & $-0.27^{+0.68}_{-0.61}$ \\
             $\beta_2$ & $(1.1\pm0.3)\times10^{-2}$ & $(5.6^{+9.2}_{-8.9})\times10^{-3}$ \\
             \hline
             $\gamma_0$ & $-1558.46^{+313.76}_{-299.25}$ & $-398.31^{+262.44}_{-236.41}$ \\
             $\gamma_1$ & $46.91^{+11.41}_{-14.15}$ & $32.33^{+8.00}_{-8.45}$ \\
             $\gamma_2$ & $27.77^{+4.93}_{-4.57}$ & $10.43^{+3.21}_{-3.70}$ \\
             \hline
        \end{tabular}
        \label{tab:planes}
\end{table}

\begin{table}
        \centering
        \caption{Values for $\beta_0$, $\beta_1$, and $\beta_2$ obtained using the results from the alpha and gamma parameterisations and rearranging Equations~\ref{eq:alpha} and \ref{eq:gamma} to Equation~\ref{eq:beta}.}
        \renewcommand{\arraystretch}{1.4}
        \begin{tabular}{lccc}
        \hline
        \hline
             Parameterisation & $\beta_0$ & $\beta_1$ & $\beta_2$ \\ 
             \hline
             \textit{Entire population} \\
             $\alpha$ & $388.40^{+811.87}_{-1244.54}$ & $-12.32^{+43.10}_{-28.24}$ & $0.02\pm0.05$ \\
             $\gamma$ & $57.20^{+15.79}_{-15.07}$ & $-1.70^{+0.66}_{-0.73}$ & $0.04\pm0.01$ \\
             \hline
             \textit{Overlapping region} \\
             $\alpha$ & $352.59^{+748.20}_{-1173.77}$ & $-10.70^{+39.12}_{-24.76}$ & $0.01\pm0.09$ \\
             $\gamma$ & $39.20^{+27.00}_{-22.99}$ & $-3.06^{+1.16}_{-2.06}$ & $0.09^{+0.05}_{-0.02}$ \\
             \hline
        \end{tabular}
        \label{tab:re}
\end{table}

\subsection{Overcoming current limitations}

The interior structure model used in this work relies on a number of simplifying assumptions, primarily, a layered interior.
Nevertheless, the results provide a meaningful relative sense of the compositional differences between the planets in our sample on a population level, even if absolute values should be interpreted with caution.

A further complexity not intentionally captured by our current models is the sensitivity of interior structure to planetary temperature.
A more fundamental limitation is the assumption of discrete, chemically distinct layers that underlies \texttt{plaNETic} and \texttt{BICEPS}, namely an iron core, a silicate mantle, a water layer, and an amount of H/He mixed into the water layer.
At the temperatures and pressures of irradiated sub-Neptune interiors, the boundaries between these reservoirs are not sharp: hydrogen and silicates become mutually soluble, magma oceans are common and may be permanent, and supercritical conditions erase the interfaces between nominally separate shells \citep{Young2024, Rogers2025b, Young2025, Calder2026, Nicholls2026b}.
Because the iron mass fraction we infer is obtained by partitioning the planetary mass into these discrete layers, enforcing a layered structure where it is not physically realised may bias the retrieved composition; the continual chemical exchange between molten interiors and their atmospheres, and the imprint that escaping envelopes leave on the underlying rock, are precisely the effects that a layered model omits \citep{Lichtenberg2021, Rogers2024, Lichtenberg2025}.
Since this miscibility is expected to be most pronounced in the hottest planets, it could also contribute to the apparent dependence on equilibrium temperature.
Overcoming this will require interior models that relax the layered assumption and follow the coupled evolution of atmosphere and interior self-consistently \citep{Rogers2025, Nicholls2026}; until such models become standard, both the absolute \fironp values and any temperature-dependent trends should be regarded as provisional.

To fully address these limitations, future work should pursue two main avenues.
First, expanding the sample to cover a wider range of equilibrium temperatures would allow trends to be characterised more robustly and help isolate the role of temperature in driving compositional variation.
Second, formulating more complex interior structure models will be essential for moving beyond the relative comparisons towards more precise constraints on planetary composition.

As we do not know the specific formation pathways for each planet, we make necessarily arbitrary assumptions regarding which set of \texttt{plaNETic} priors to use (water-poor or water-rich).
Since the formation location of a planet relative to its system's ice-line determines whether it accretes water-rich or water-poor material, the choice of priors carries physical implications.
Use of water-poor priors implies planetary formation within a system's ice-line, whereas water-rich priors are more appropriate for planets that formed beyond it and subsequently migrated inwards.
In the absence of evidence to favour either scenario, we adopted water-rich priors as default, acknowledging that this choice introduces a source of uncertainty in the retrieved compositions.

A further limitation concerns the physical interpretation of any dependence on equilibrium temperature.
$T_\mathrm{eq}$ is a present-day quantity, set by the stellar irradiation a planet currently receives, and it is a poor proxy for the disc environment in which the planet's material was assembled and chemically processed.
A planet's solids are sourced from a range of disc locations and growth stages, and subsequent migration further decouples the present orbit from the formation site \citep{Drazkowska2023}.
The compositional memory of formation, moreover, is carried predominantly by the volatile budget and oxidation state rather than by the refractory iron-to-silicate ratio; this budget is governed by disc chemistry, the location of ice lines, and the thermal and collisional processing of the building blocks during accretion, and can be altered substantially at successive stages of planet formation \citep{Krijt2023, Kong2026}.
The processes that do change the bulk iron fraction, such as the preferential stripping of silicate mantles in high-energy collisions that leaves iron-enriched remnants, are collisional in origin and likewise decoupled from a planet's equilibrium temperature \citep{Asphaug2014}.
A correlation between $T_\mathrm{eq}$ and planetary iron content would therefore be more naturally attributed to post-formation processing of the planet itself, such as the escape and devolatilisation discussed above, than to an inherited formation-temperature gradient.

\section{Conclusions}\label{sec:conc}

In this work, we investigated the relationship between stellar and planetary composition for super-Earths and sub-Neptunes, employing a four-component treatment of planetary interiors; this provides a deeper insight into the compositions of exoplanets than simpler models permit.
Our results suggest, however, that with the current sample and compositional proxies used, this relationship cannot be modelled with sufficient precision or accuracy to draw robust conclusions.
We caution that uncertainties on stellar abundances are likely underestimated in many other cases, further limiting the reliability of any inferred trends.

We find trends with equilibrium temperature in our results; however, we caution that this should not be over-interpreted, as assumptions within the planetary interior models used in the study may unintentionally introduce or amplify such a trend.
Disentangling a genuine dependence between planetary composition and equilibrium temperature from modelling artefacts will therefore require careful consideration in future work.

Furthermore, one of the main limitations of this study is the sample itself, both in terms of size and also coverage in various parameter spaces (such as equilibrium temperature).
To draw robust conclusions about the stellar–planetary compositional relationship, a larger catalogue of precisely characterised systems is essential, particularly super-Earths, which remain under-represented in well-constrained datasets.

Despite these limitations, this study demonstrates the value of more physically detailed treatments of planetary structure.
As stellar atmospheric models improve, sample sizes increase, and planetary interior structure models are refined, studies of this kind will be better-positioned to uncover the true nature of the compositional link between stars and their planets.

\section*{Acknowledgements}

We would like to thank the referee for their constructive and prompt report.

We would also like to thank Manfredo Capriolo and Stephen Jones for their assistance and useful discussions.

The spectra in this paper were collected under the following programme IDs:

\textbf{ESPRESSO:} 104.20U8.001, 105.20P7.001, 106.218R.001, 106.21M2.002, 106.21M2.003, 106.21M2.004, 106.21M2.007, 108.2254.002, 108.2254.003, 108.2254.006, 108.22GM.001, 110.24CD.002, 110.24CD.003, 110.24CD.009, 1102.C-0744, 1102.C-0958, 1104.C-0350

\textbf{HARPS:} 0101.C-0510, 0101.C-0829, 0102.C-0525, 072.C-0488, 082.C-0308, 088.C-0323, 106.21ER.001, 106.21TJ.001, 108.22KV.001, 108.22KV.002, 108.22KV.003, 108.22KV.005, 1102.C-0249, 1102.C-0923, 112.25WB.002, 183.C-0972, 198.C-0169, 282.C-5036, 60.A-9700

\textbf{HARPS-N:} A32DDT2, A36TAC\_12, A40TAC\_23, A43DDT2, A48TAC\_59, CAT15B\_92, CAT16B\_61, CAT18A\_115, CAT18A\_34, CAT18B\_62, CAT19A\_159, CAT19A\_162, CAT19A\_96, CAT19B\_154, CAT20B\_80, CAT21A\_119, GAPS, GTO, ITP15\_7, ITP19\_1, OPT13B\_30, OPT17B\_59

\textbf{SOPHIE:} 11A.PNP.CONS, 12B.PNP.CONS, 19A.PNP.HEBR

DAT acknowledges the support of the Science and Technology Facilities Council (STFC).

AM acknowledges a UK Science and Technology Facilities Council (STFC) small grant ST/Y002334/1 and funding from a UKRI Future Leader Fellowship, grant number MR/X033244/1.

TL was supported by the Branco Weiss Foundation, the European Research Council (ERC) under the European Union's Horizon Europe research and innovation programme (MagmaWorlds, 101219807), the Alfred P. Sloan Foundation (AEThER, G-2025-25284), NASA’s Nexus for Exoplanet System Science research coordination network (Alien Earths, 80NSSC21K0593), and the NWO NWA-ORC PRELIFE Consortium (NWA.1630.23.013).

AAAP acknowledges funding from a UK Science and Technology Facilities Council (STFC) Small Award, grant number UKRI/ST/B001171/1.


\section*{Data Availability}

Data will be available in an online table on Vizier CDS.



\bibliographystyle{mnras}
\bibliography{bib} 



\appendix

\section{Spectrographs used}

Table \ref{tab:instruments} lists the spectrographs that were used for each star to measure the stellar atmospheric parameters and elemental abundances.

\begin{table}
        \centering
        \caption{Instruments from which archival spectra were drawn for each target and the S/N of the co-added spectra used.}
        \renewcommand{\arraystretch}{1.2}
        \begin{tabular}{llp{0.33\columnwidth}}
        \hline
        \hline
            Target & Instrument & S/N \\ 
            \hline
            55 Cnc & HARPS-N & 413 \\
            BD+38 5036 & HARPS-N & 484 \\
            BD+39 2643 & HARPS-N & 488 \\
            CoRoT-7 & HARPS & 556 \\
            EPIC 210894022 & ESPRESSO & 646 \\
            EPIC 211682544 & HARPS-N & 332 \\
            EPIC 212779563 & HARPS-N & 326 \\
            EPIC 220709978 & HARPS-N & 594 \\
            EPIC 229004835 & HARPS-N & 1528 \\
            HD 136352 & HARPS & 535 \\
            HD 207897 & SOPHIE & 648 \\
            HD 212657 & HARPS-N & 464 \\
            HD 213885 & HARPS & 825 \\
            HD 3167 & ESPRESSO & 490 \\
            HIP 9618 & SOPHIE & 237 \\
            K2-131 & HARPS-N & 220 \\
            K2-138 & HARPS & 488 \\
            K2-291 & HARPS-N & 323 \\
            Kepler-107 & HARPS-N & 341 \\
            Kepler-19 & HARPS-N & 382 \\
            Kepler-20 & HARPS-N & 375 \\
            Kepler-93 & HARPS-N & 995 \\
            TOI-1062 & HARPS & 518 \\
            TOI-1064 & ESPRESSO & 357 \\
            TOI-1246 & HARPS-N & 332 \\
            TOI-125 & HARPS & 528 \\
            TOI-1339 & HARPS-N & 476 \\
            TOI-1416 & HARPS-N & 416 \\
            TOI-1469 & HARPS-N & 349 \\
            TOI-2000 & HARPS & 363 \\
            TOI-220 & HARPS & 531 \\
            TOI-238 & ESPRESSO & 811 \\
            TOI-2458 & HARPS & 362 \\
            TOI-266 & ESPRESSO & 851 \\
            TOI-286 & ESPRESSO & 890 \\
            TOI-402\textsuperscript{\textdagger} & ESPRESSO & 1239 \\
            TOI-469 & ESPRESSO & 803 \\
            TOI-509 & HARPS-N & 395 \\
            TOI-560 & HARPS & 590 \\
            TOI-561 & HARPS-N & 911 \\
            TOI-652 & HARPS & 522 \\
            TOI-733 & HARPS & 558 \\
            TOI-755 & HARPS & 561 \\
            TOI-763 & HARPS & 557 \\
            WASP-47 & HARPS-N & 365 \\
            \hline
            \multicolumn{3}{@{}p{0.8\columnwidth}@{}}{\small \textsuperscript{\textdagger}ESPRESSO medium resolution mode was used for TOI-402 ($\mathcal{R}=70\,000$).} \\
        \end{tabular}
        \label{tab:instruments}
\end{table}

\section{New vs archival stellar properties}
\label{sec:new_vs_arch}

\begin{figure*}
	\includegraphics[width=0.85\textwidth]{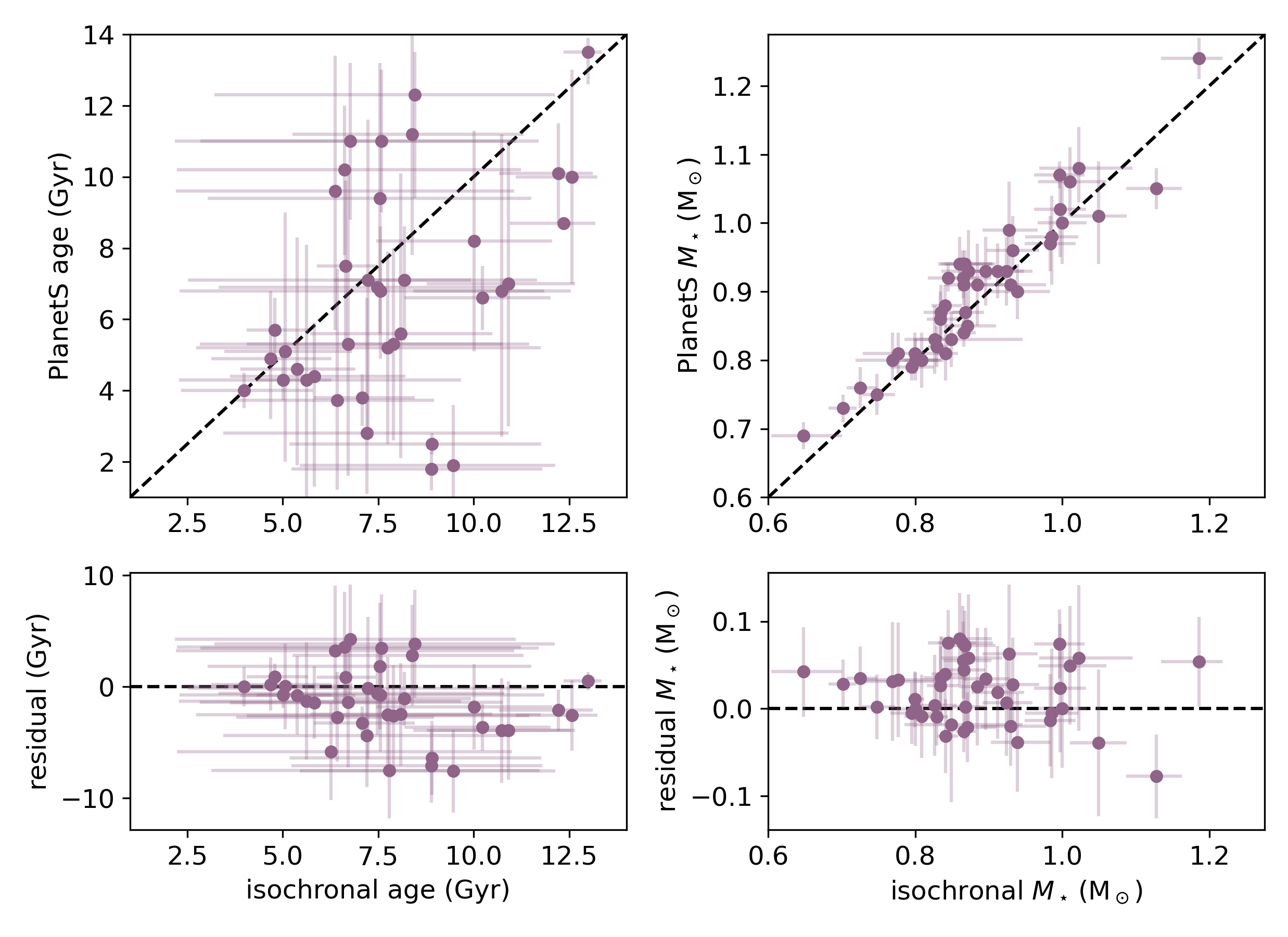}
    \vspace{-0.5cm}
    \includegraphics[width=0.85\textwidth]{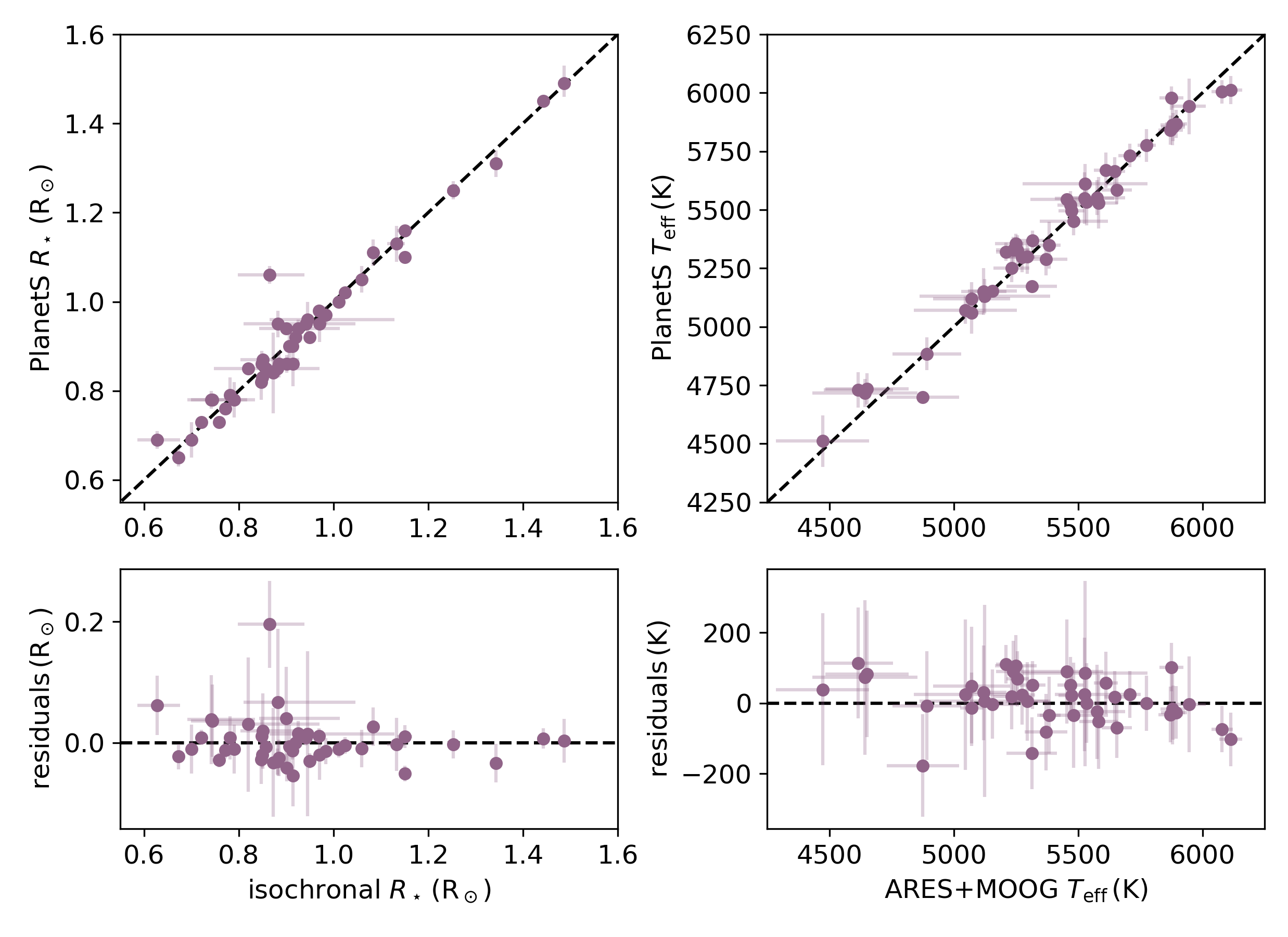}
    \caption{Comparison between the values calculated in this work and those from PlanetS for age, $M_\star$, $R_\star$, and $T_\mathrm{eff}$.}
    \label{fig:new_vs_arch}
\end{figure*}

Fig. \ref{fig:new_vs_arch} compares some stellar parameters measured by this work versus those used in the PlanetS catalogue. 

\section{Plane fits}\label{sec:plane}

\begin{figure}
	\includegraphics[width=0.99\columnwidth, trim={2cm 0 1.5cm 0}, clip]{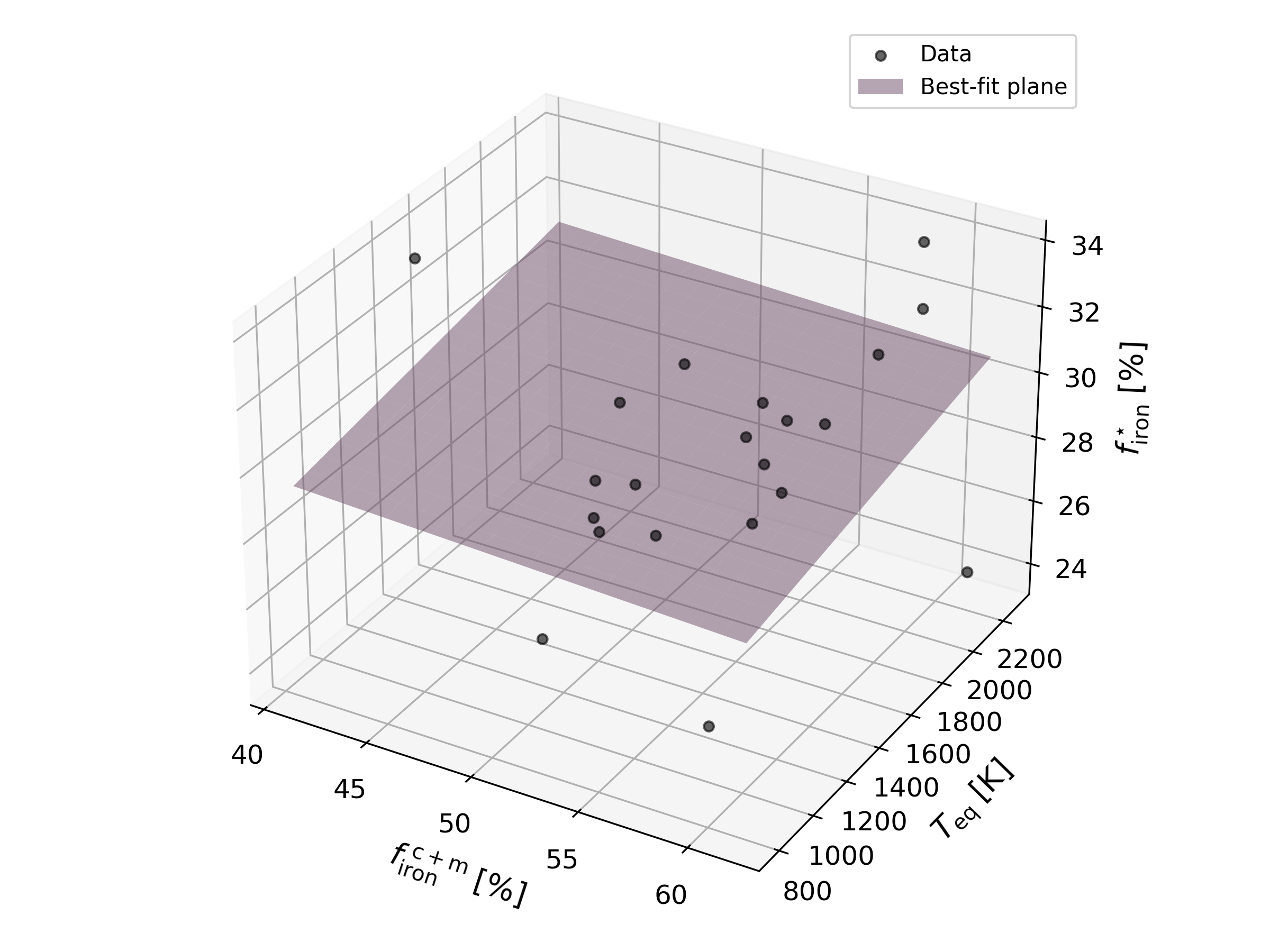}
    \caption{Plane fit to the super-Earths in the sample using the alpha parameterisation (Equation~\ref{eq:alpha}).}
    \label{fig:sE_alpha}
\end{figure}

\begin{figure}
	\includegraphics[width=0.99\columnwidth, trim={2cm 0 1.5cm 0}, clip]{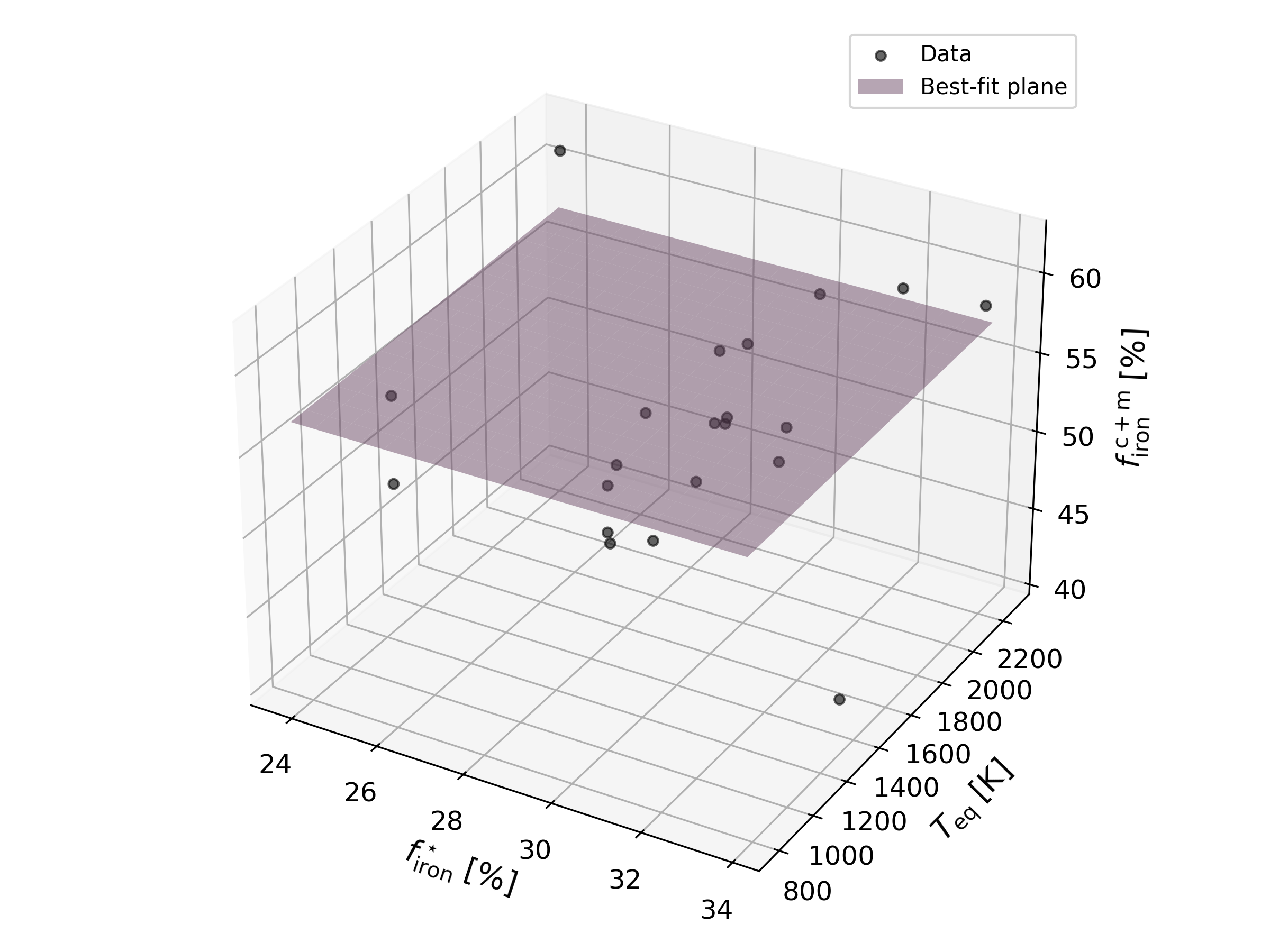}
    \caption{Plane fit to the super-Earths in the sample using the beta parameterisation (Equation~\ref{eq:beta}).}
    \label{fig:sE_beta}
\end{figure}

\begin{figure}
	\includegraphics[width=0.99\columnwidth, trim={2cm 0 1.5cm 0}, clip]{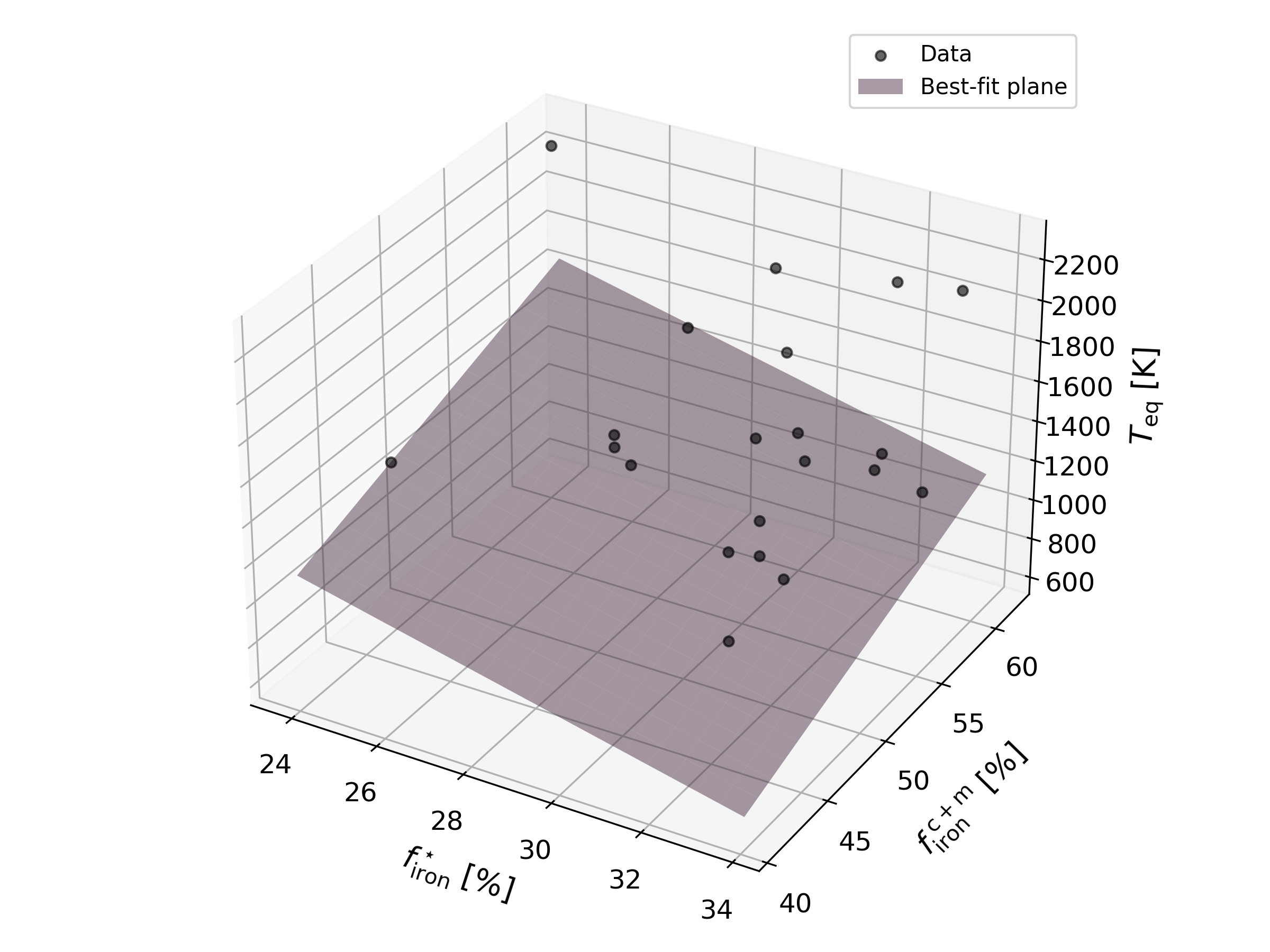}
    \caption{Plane fit to the super-Earths in the sample using the gamma parameterisation (Equation~\ref{eq:gamma}).}
    \label{fig:sE_gamma}
\end{figure}

\begin{figure}
	\includegraphics[width=0.99\columnwidth, trim={2cm 0 1.5cm 0}, clip]{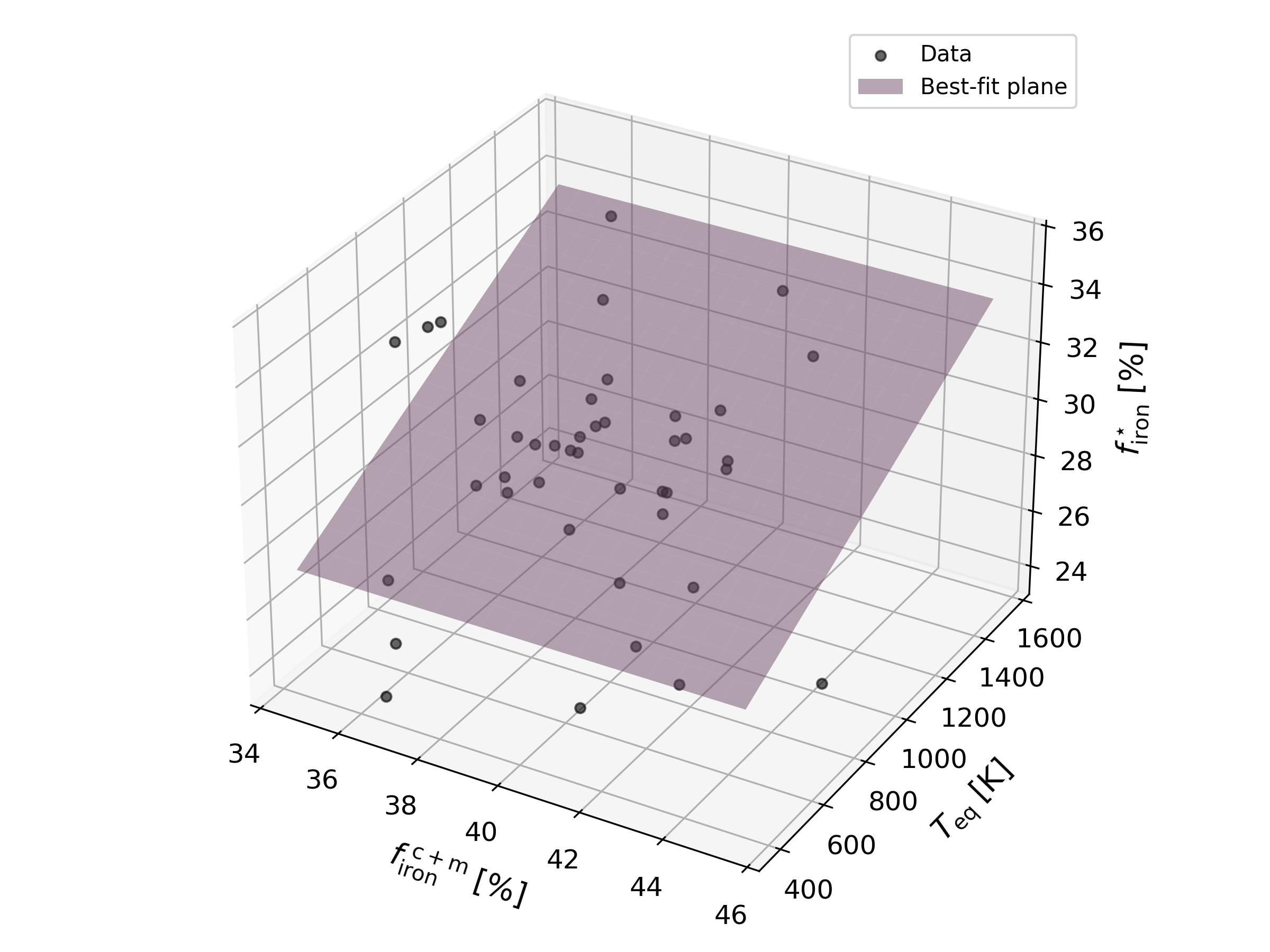}
    \caption{Plane fit to the sub-Neptunes in the sample using the alpha parameterisation (Equation~\ref{eq:alpha}).}
    \label{fig:sN_alpha}
\end{figure}

\begin{figure}
	\includegraphics[width=0.99\columnwidth, trim={2cm 0 1.5cm 0}, clip]{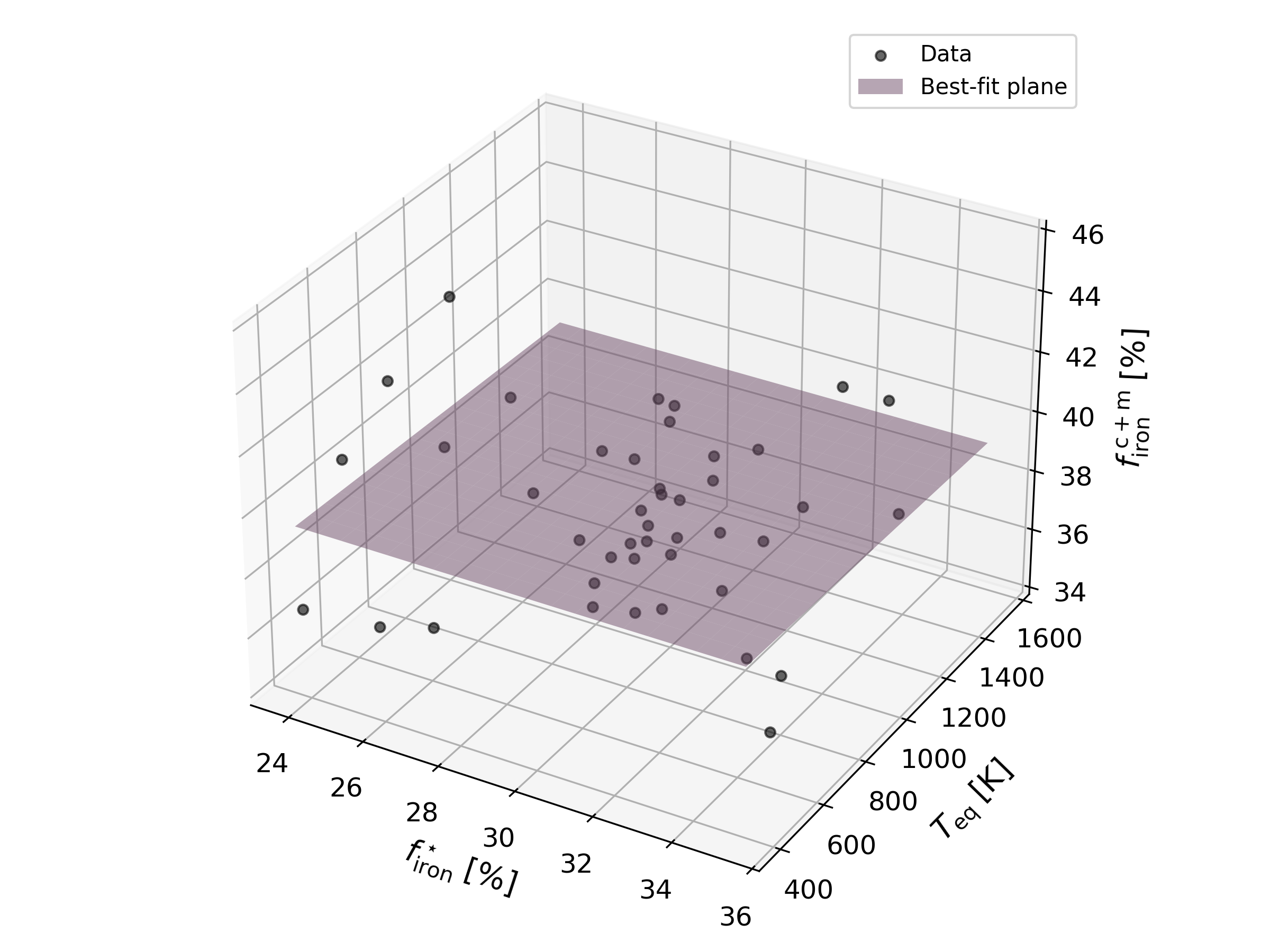}
    \caption{Plane fit to the sub-Neptunes in the sample using the beta parameterisation (Equation~\ref{eq:beta}).}
    \label{fig:sN_beta}
\end{figure}

\begin{figure}
	\includegraphics[width=0.99\columnwidth, trim={2cm 0 1.5cm 0}, clip]{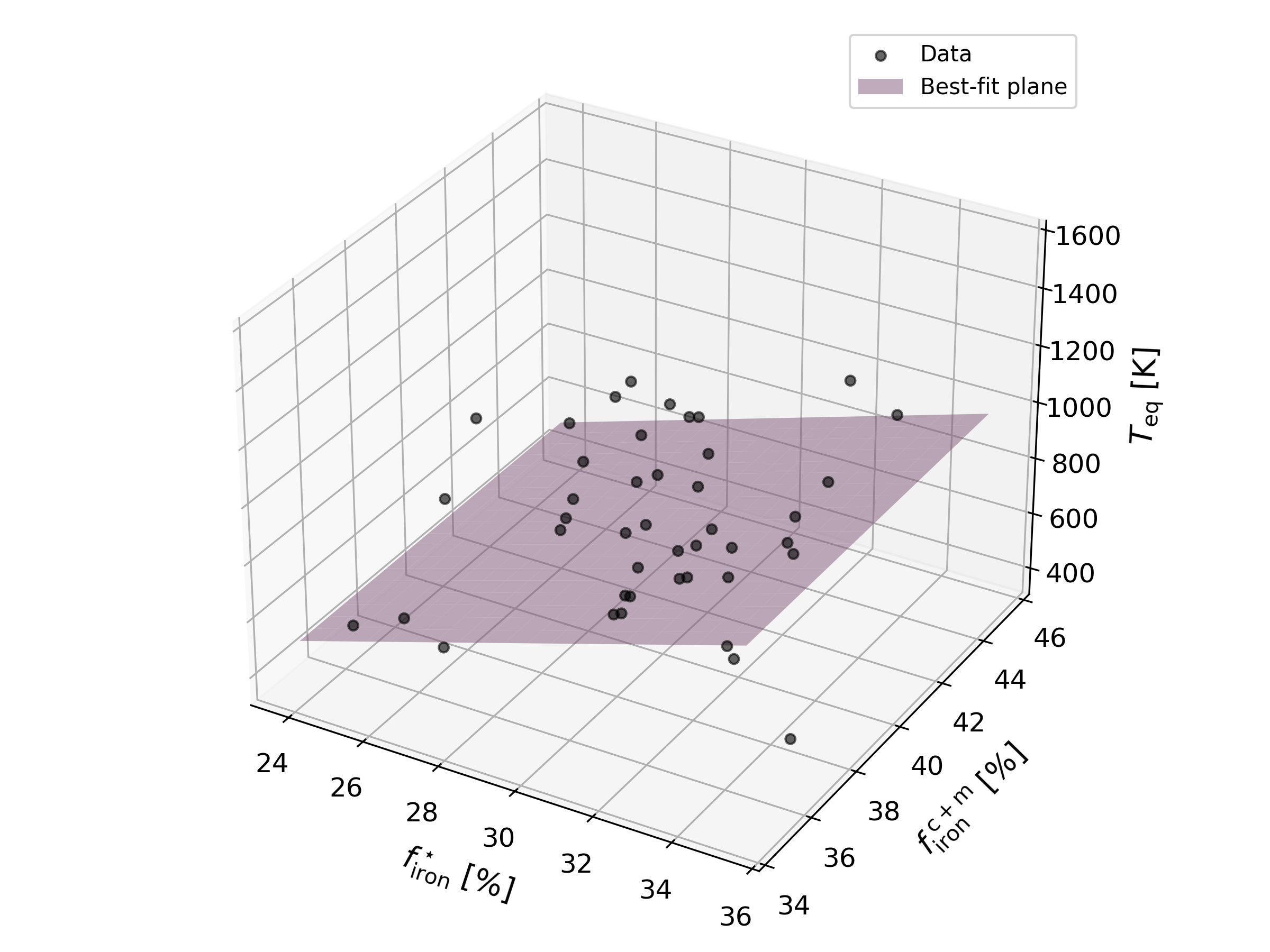}
    \caption{Plane fit to the sub-Neptunes in the sample using the gamma parameterisation (Equation~\ref{eq:gamma}).}
    \label{fig:sN_gamma}
\end{figure}

\begin{figure}
	\includegraphics[width=0.99\columnwidth, trim={2cm 0 1.5cm 0}, clip]{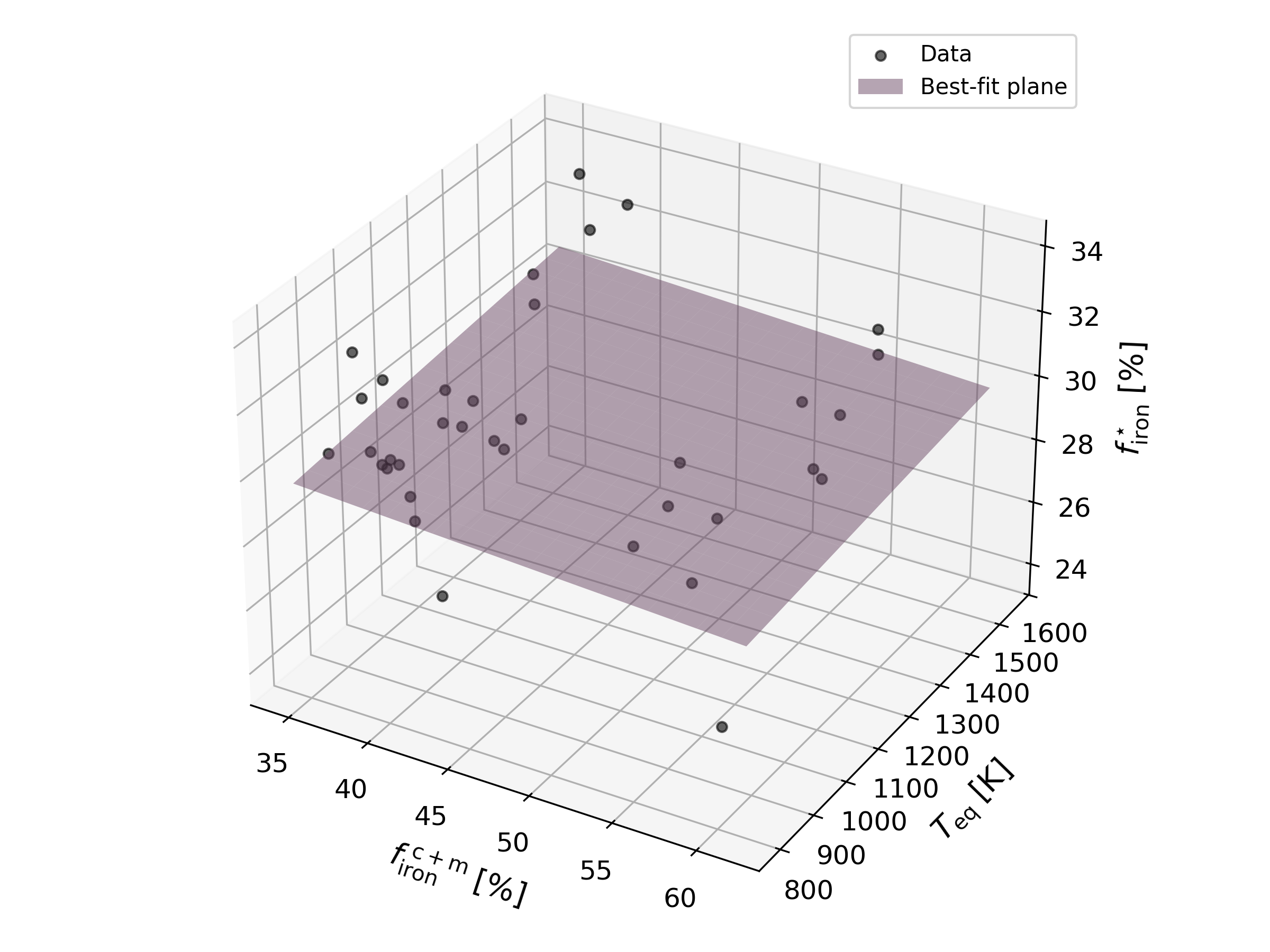}
    \caption{Plane fit to the overlapping region of the sample using the alpha parameterisation (Equation~\ref{eq:alpha}).}
    \label{fig:overlap_alpha}
\end{figure}

\begin{figure}
	\includegraphics[width=0.99\columnwidth, trim={2cm 0 1.5cm 0}, clip]{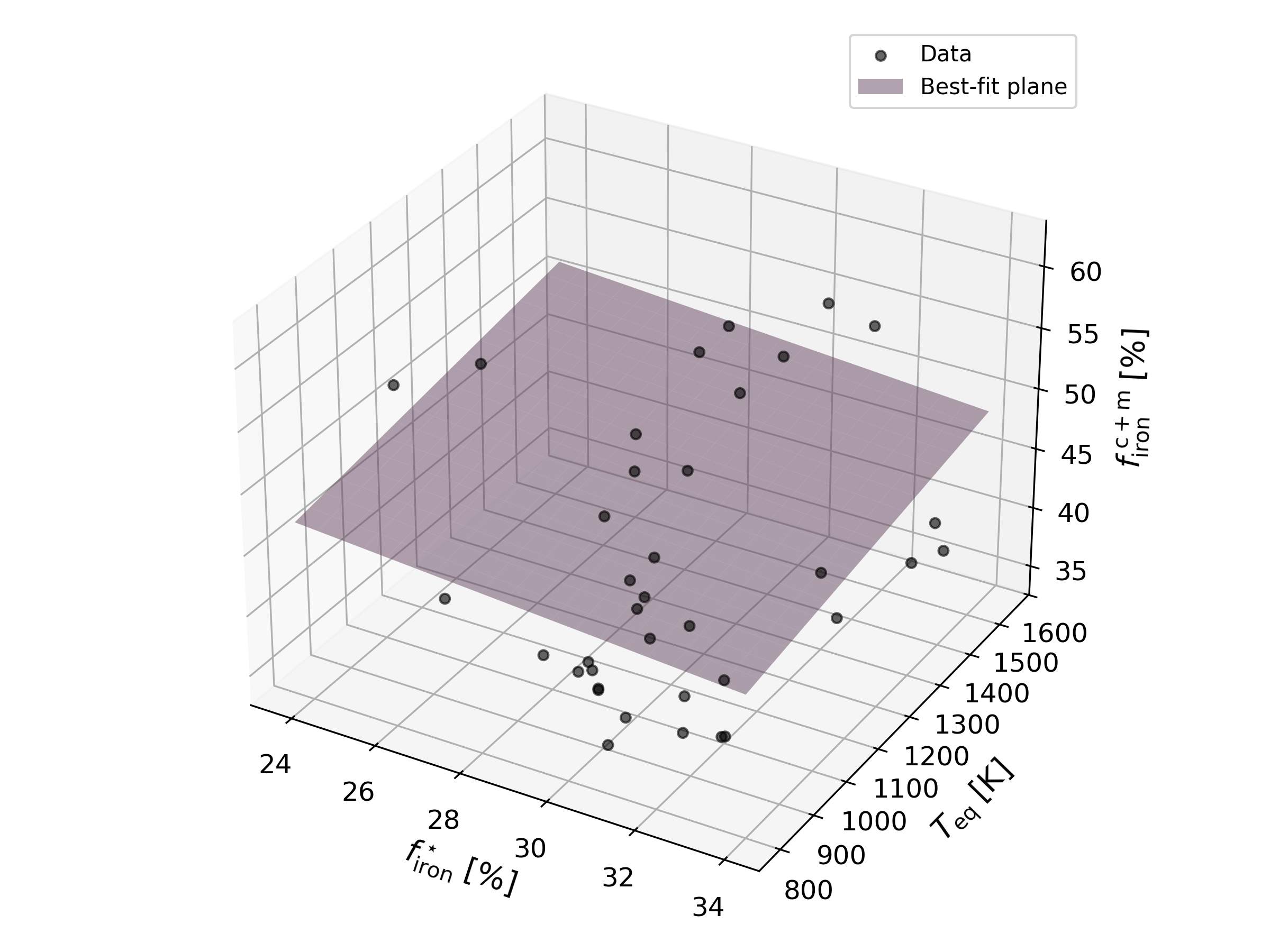}
    \caption{Plane fit to the overlapping region of the sample using the beta parameterisation (Equation~\ref{eq:beta}).}
    \label{fig:overlap_beta}
\end{figure}

\begin{figure}
	\includegraphics[width=0.99\columnwidth, trim={2cm 0 1.5cm 0}, clip]{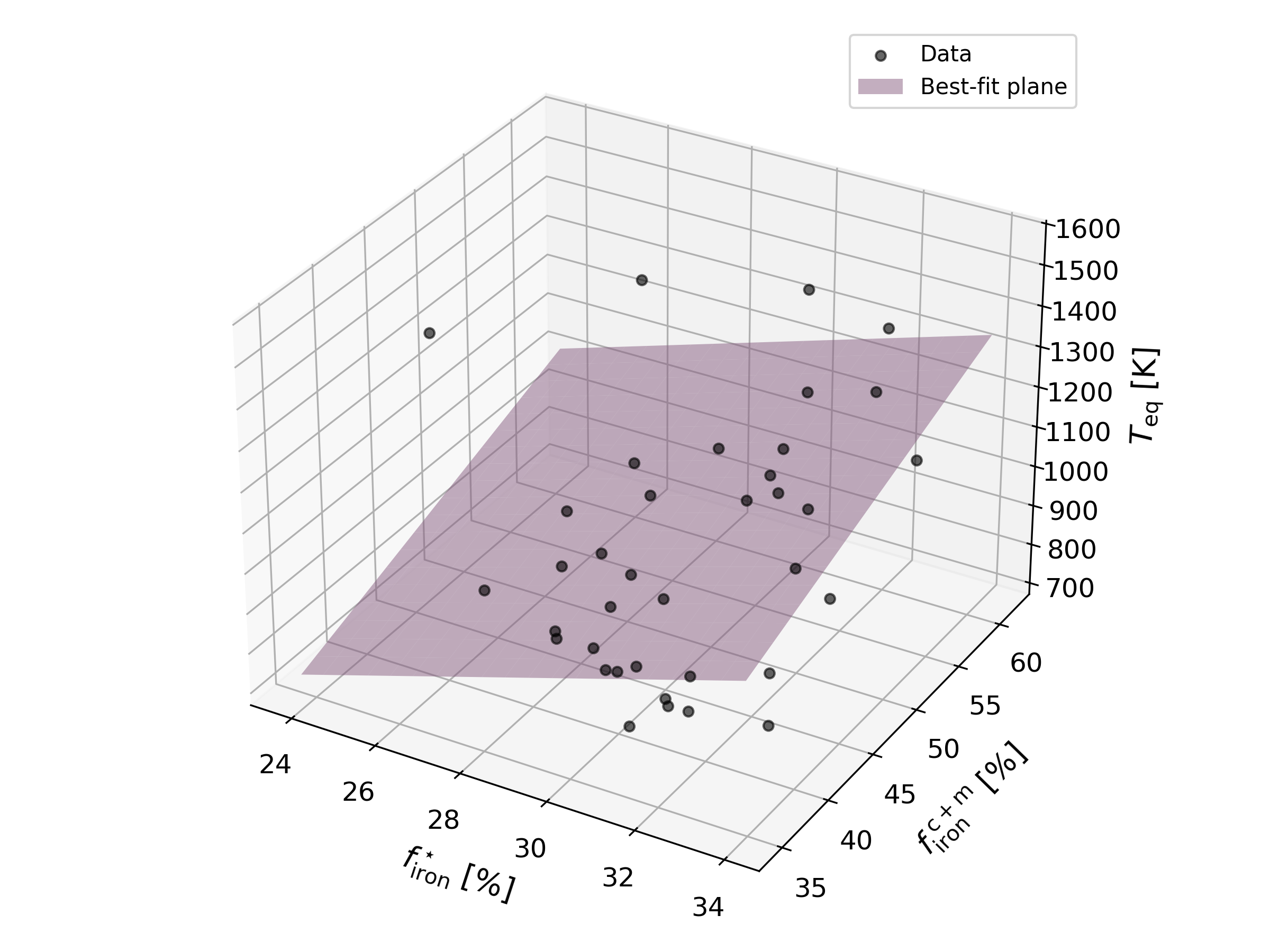}
    \caption{Plane fit to the overlapping region of the sample using the gamma parameterisation (Equation~\ref{eq:gamma}).}
    \label{fig:overlap_gamma}
\end{figure}

Table \ref{tab:sub_planes} lists all fitted coefficients for the three different parameterisations of the plane linking planet iron mass fraction, stellar iron mass fraction and equilibrium temperature for the super-Earths and sub-Neptunes. Figs. \ref{fig:sE_alpha}--\ref{fig:overlap_gamma} show the fitted planes for all parameterisations for the super-Earth, sub-Neptune and overlap sample. 

\begin{table}
        \centering
        \caption{Best fit values for the three plane parameterisations (super-Earths and sub-Neptunes).}
        \renewcommand{\arraystretch}{1.4}
        \begin{tabular}{lcc}
        \hline
        \hline
             Parameters & Super-Earths & Sub-Neptunes \\ 
             \hline
             $\alpha_0$ & $29.78^{+9.07}_{-8.59}$ & $25.48^{+4.00}_{-3.87}$ \\
             $\alpha_1$ & $(-0.3^{+2.6}_{-2.5})\times 10^{-3}$ & $(5.4\pm3.1)\times 10^{-3}$ \\
             $\alpha_2$ & $(-0.07^{+14.46}_{-14.74})\times 10^{-2}$ & $(-0.11\pm7.6)\times 10^{-2}$ \\
             \hline
             $\beta_0$ & $56.55\pm13.92$ & $39.75^{+11.35}_{-11.38}$ \\
             $\beta_1$ & $(-0.23^{+39.37}_{-40.21})\times 10^{-2}$ & $(-0.64^{+34.43}_{-34.41})\times 10^{-2}$ \\
             $\beta_2$ & $(0.4^{+4.1}_{-4.0})\times 10^{-3}$ & $(-0.2^{+7.7}_{-7.9})\times 10^{-3}$ \\
             \hline
             $\gamma_0$ & $1259.52^{+3619.35}_{-3380.68}$ & $-475.03^{+517.88}_{-277.29}$ \\
             $\gamma_1$ & $-49.63^{+122.70}_{-37.96}$ & $40.66^{+6.24}_{-8.16}$ \\
             $\gamma_2$ & $25.71^{+25.33}_{-56.88}$ & $0.48^{+9.13}_{-9.86}$ \\
             \hline
        \end{tabular}
        \label{tab:sub_planes}
\end{table}

\bsp	
\label{lastpage}
\end{document}